\UseRawInputEncoding
\documentclass[journal]{IEEEtran}
\usepackage{multirow}
\usepackage{diagbox}
\usepackage{amssymb}
\usepackage{amsmath}

\usepackage{graphicx}
\usepackage{cite}
\usepackage{citesort}
\usepackage{subfigure}
\usepackage{graphicx,epstopdf}
\usepackage{epsfig}	
\usepackage{cite,graphicx,amsmath,amssymb}
\usepackage{comment}

\usepackage{amssymb}
\usepackage{amsmath}
\usepackage{cite}
\usepackage{url}
\usepackage{xcolor}
\usepackage{cite,graphicx,amsmath,amssymb}
\usepackage{subfigure}
\usepackage{citesort}
\usepackage{fancyhdr}
\usepackage{mdwmath}
\usepackage{mdwtab}
\usepackage{caption}
\usepackage{amsthm}
\usepackage{lipsum}

\newtheorem{theorem}{Theorem}

\newtheorem{lemma}{Lemma}

\newtheorem{corollary}{Corollary}

\newtheorem{remark}{Remark}  

\makeatletter
\def\ScaleIfNeeded{%
\ifdim\Gin@nat@width>\linewidth \linewidth \else \Gin@nat@width
\fi } \makeatother

\begin{document}

\title{\Huge{Performance Analysis of Reconfigurable Intelligent Surface Assisted Two-Way NOMA Networks}}

\author{Ziwei~Liu, Xinwei~Yue,~\IEEEmembership{Senior Member IEEE,} Chao Zhang,~\IEEEmembership{Student Member,~IEEE,}  Yuanwei Liu,~\IEEEmembership{Senior Member,~IEEE,}  Yuanyuan Yao, Yafei~Wang and Zhiguo Ding,~\IEEEmembership{Fellow,~IEEE}


\thanks{Z. Liu,  X. Yue, Y. Yao and Y. Wang are with the Key Laboratory of Modern Measurement $\&$ Control Technology, Ministry of Education and also with the School of Information and Communication Engineering, Beijing Information Science and Technology University, Beijing 100101, China. (email: \{ziwei.liu, xinwei.yue, yyyao and wangyafei\}@bistu.edu.cn).}
\thanks{C. Zhang and Y. Liu are with the School of Electronic Engineering and Computer Science, Queen Mary University of London, London E1 4NS, U.K. (email: \{chao.zhang and yuanwei.liu\}@qmul.ac.uk).}
\thanks{Z. Ding is with the Department of Electrical Engineering, Princeton University, Princeton, USA and also with the School of Electrical and Electronic Engineering, the University of Manchester, Manchester, U.K (e-mail: zhiguo.ding@manchester.ac.uk).}
}

\maketitle
\begin{abstract}
This paper investigates the performance of reconfigurable intelligent surface assisted two-way non-orthogonal multiple access (RIS-TW-NOMA) networks, where a pair of users exchange their information through a RIS. The influence of imperfect successive interference cancellation on RIS-TW-NOMA is taken into account. To evaluate the potential performance of RIS-TW-NOMA, we derive the exact and asymptotic expressions of outage probability and ergodic rate for a pair of users. Based on the analytical results, the diversity orders and high signal-to-noise ratio (SNR) slopes are obtained in the high SNR regime, which are closely related to the number of RIS elements. Additionally, we analyze the system throughput and energy efficiency of RIS-TW-NOMA networks in both delay-limited and delay-tolerant transmission modes. Numerical results indicate that: 1) The outage behaviors and ergodic rate of RIS-TW-NOMA are superior to that of RIS-TW-OMA and two-way relay OMA (TWR-OMA); 2) As the number of RIS elements increases, the RIS-TW-NOMA networks are capable of achieving the enhanced outage performance; and 3) By comparing with RIS-TW-OMA and TWR-OMA networks, the energy efficiency and system throughput of RIS-TW-NOMA has obvious advantages.
\end{abstract}

\begin{keywords}
Reconfigurable intelligent surface, two way non-orthogonal multiple access,  imperfect SIC, outage probability, ergodic rate.
\end{keywords}

\section{Introduction}
With the rapid advancement in the wireless communication networks, reconfigurable intelligent surface (RIS) has been considered as one of promising techniques to improve the spectrum efficiency for the sixth-generation (6G) communications  \cite{9086766,9144301,9424177}. The main feature of RIS is to reconfigure the incident signals by the virtue of a programmable controller in full-duplex mode, which not requires the enhanced self-interference cancellation techniques. More specifically, the RIS assisted wireless communications were discussed in \cite{8796365,9060923}, where the signal quality of receivers are boosted and the channel interference are suppressed by adjusting the reflection amplitude or phase of each passive reflecting element.
As such, RIS have a great potential to revolutionize the design of wireless networks.
In \cite{hou2020reconfigurable}, some typical applications of RIS-aided wireless communications were introduced to create the virtual line-of-sight (LoS) links and extend the coverage extension.

Compared to the conventional relay networks,  the RIS is capable of controlling the signal without the needs of complex decoding and encoding \cite{Gong2020IRS,Kudathanthirige2020RIS,2020throughputIRS}. To exploit the performance gain, the authors of \cite{Gong2020IRS} discussed the rate maximization and RIS's phase control problems for RIS-assisted wireless networks. In \cite{Kudathanthirige2020RIS}, the symbol error probability and achievable rate were investigated of RIS intended for aiding wireless communications. Furthermore, the authors of \cite{2020throughputIRS} considered a multi-RIS-assisted system, which deploys different strategies for RIS to maximize the spatial throughput. With the objective of improving energy efficiency, the energy-efficient strategy of RIS-assisted multi-user systems was designed  in \cite{Huang2019RISEE} by premeditating both the transmit power allocation and phase shifting of reflecting elements. In \cite{Diluka2020RIS}, the outage probability of a distributed RIS-aided communication system was studied over Nakagami-$m$ fading channels. As a further advance, the authors of \cite{9146875} researched the ergodic capacity of RIS-aided communication networks over Rician fading channels.
From the perspective of practical issues, the authors of \cite{Jiayi2021RIS} analyzed the coverage probability of RIS-aided communication networks.
Lately, the concept of simultaneously transmitting and reflecting (STAR)-RIS was proposed in \cite{Liu2021STAR360},  where the incident signals can not
only be reflected in the same side of RIS, but also can be transmitted to the other side of RIS.

The above treatises mainly focus attention on the RIS-assisted conventional orthogonal multiple access (OMA) communication systems. In actual, non-orthogonal multiple access (NOMA) has been surveyed at large, which is capable of improving the system efficiency and the number of users connected \cite{Ding2014Randomly,Liu8114722,2016Nonorthogonal}. The ideology of cooperative NOMA communications was proposed in \cite{Ding2015Cooperative}, where the nearby users are selected to be relays to forward the superposed signals to the distant users. A cooperative simultaneous wireless power transfer aided NOMA protocol was proposed in \cite{liu2016cooperative} , where a NOMA user benefitting from good channel conditions acts as an energy harvesting source in order to assist a NOMA user suffering from poor channel conditions.
Furthermore, the authors of \cite{Yue2018NOMA} evaluated the outage behaviors and ergodic rate of  full-duplex (FD) cooperative NOMA systems.
On the other hand, a class of dedicated relay assisted cooperative NOMA schemes were developed, where the relay adopts either  amplify-and-forward (AF) or decode-and-forward (DF) protocol. In \cite{Men2017NOMANakagami}, the closed-form expressions of the outage probability and ergodic sum rate were derived for AF relay-assisted NOMA networks. To enhance cell edge coverage, the authors of \cite{8704259} discussed the application of selective and incremental-selective DF relays to NOMA networks. In addition,  a novel hybrid power allocation strategy was introduced to DF relay-assisted NOMA networks \cite{Wan8353359}, which  can reduce the signaling overhead at the expense of marginal sum rate degradation.

Until now, many contributions have applied the RIS to assist NOMA communications in \cite{111,ZhengIRSUserpairing,9345507,Ding2019IRS}. More specifically, the authors of \cite{111} investigated the rate performance of RIS-assisted NOMA networks by jointly designing the power allocation of base station and phase shifts at the RIS. In \cite{ZhengIRSUserpairing}, the theoretical behaviors of RIS-NOMA were compared to traditional NOMA without RIS and OMA with/without RIS.  By considering downlink and uplink communications, the performance of RIS-NOMA was characterized in terms of the outage probability and ergodic rate over Nakagami-$m$ fading channels \cite{9345507}.  Explicit insights for understanding the  co-existence of RIS and NOMA, a simple design of RIS-NOMA was proposed in \cite{Ding2019IRS}, where the outage probability of non-orthogonal users are derived with finite resolution beamforming. As a further advance, the authors of  \cite{9079918} investigated the outage probability of  RIS-NOMA networks by employing coherent phase and random phase shifting. With the help of 1-bit coding scheme, the authors in \cite{Yue2021NOMARIS} investigated the outage probability, ergodic rate and energy efficiency of RIS-NOMA networks, in which both imperfect successive interference cancellation (ipSIC) and perfect successive interference cancellation (pSIC) are taken into consideration. In \cite{2020LargeNOMARIS}, the pairwise error probability of multiple users for RIS-NOMA networks were evaluated under the assumption of ipSIC. Recently, the outage behaviors and ergodic sum rate of STAR-RIS aided NOMA networks was surveyed in \cite{Yue2021STARNOMA,Chao2021STARNOMA} over the cascade Rician fading channels.

Above existing contributions about NOMA and RIS are discussed in one-way transmission, where the information are delivered from the base station (BS) to relay or RIS and then to terminals. As a further potential development, two-way relay (TWR) communications stated in \cite{Shannon1961Two,5403557} has sparked more attention since it is able to improve the spectrum utilization. The basic principle of TWR communications is to exchange information between a pair of nodes with the assistance of relays.
TWR-OMA users are constrained by the interference power, limiting their transmitting power. It is one of the main limiting factors to the achievable capacity of the TWR-OMA users and, consequently, the communication quality \cite{kusaladharma2016underlay}. To shed light on the TWR of NOMA networks, the authors of \cite{8120655} analyzed the TWR transmission in cooperative NOMA networks to enhance the spectral efficiency, where the TWR-NOMA networks were capable of achieving a better sum rate compared to the OMA scheme. In \cite{Yue2018TWRNOMA}, the outage probability and ergodic rate of a pair of non-orthogonal users were studied for TWR-NOMA networks. Furthermore, a network coding-based TWR-NOMA system was designed in \cite{2019TWRNOMA}, which has ability to enhance the system capacity relative to the TWR-OMA system.
As an enhancement of conventional TWR,  the few works focus on the RIS assisted two-way (TW) communications, where the relay is instead of RIS to assist the source node in sending the information. RIS is envisioned to further improve the TW-OMA link's performance by passive beamforming, resolving the performance degradation of TW-OMA networks, and opening new horizons for two-way integration communication in various new ways applications and use cases \cite{basar2019wireless}.
In \cite{Atapattu2020TWRIS}, the weighted sum rate of RIS-enhanced device-to-device OMA communications was revealed by designing the power allocation and discrete phase shifting. In addition, the authors of  \cite{Wang2021TWRIS} highlighted the performance of  RIS-assisted TWR networks to maximize SNR from the perspective of optimization.


\subsection{Motivations and Contributions}
The aforementioned research contributions have laid a solid foundation with providing a good understanding of  the RIS assisted NOMA networks, while it is still in their infancy to survey the potential benefits of TW-NOMA networks by integrating these two promising technologies in the TW networks. In view of the recent researches on 6G communication networks, RIS has been served as a potential technology for further performance improvement in TW communication networks \cite{Atapattu2020TWRIS,Wang2021TWRIS}. Hence the application of RIS for TW-NOMA networks is expected to boost the spectrum efficiency and overcome the disadvantages of traditional relays.  In addition, the use of SIC scheme still exists several potential implementation issues, i.e., error propagation and quantization errors. The ipSIC scheme is usually selected to evaluate the system performance from the practical perspective.
Motivated by the above explanations, this paper investigates RIS-TW-NOMA networks' outage probability and ergodic rate by considering ipSIC and pSIC. As a further development, the system throughput and energy efficiency of RIS-TW-NOMA with ipSIC/pSIC are discussed in both delay-limited and delay-tolerant modes.  According to the above explanations, the contributions of this paper are summarised as follows:

\begin{enumerate}
  \item
  We derive the exact expressions of outage probability for the nearby user with ipSIC/pSIC in RIS-TW-NOMA networks. We additionally derive the upper bounds and asymptotic expressions of outage probability in high SNR region. Based on the approximated results, we obtain the diversity orders of nearby users with ipSIC/pSIC schemes. We show that the outage probability of the nearby user with ipSIC performs as an error floor in the high SNR regime. 
  \item
  We also derive the exact and asymptotic expressions of outage probability for the distant user in RIS-TW-NOMA networks. To gain the corresponding diversity order, we derive the approximated outage probability of the distant user in the high SNR region. We confirm that the diversity order of the distant user is proportional to the number of RIS elements. 
   \item
   We derive the exact and approximated expressions of ergodic rate for RIS-TW-NOMA networks with ipSIC/pSIC.
   To get more insights, we further derive the high SNR slopes of a pair of users. We observe that the ergodic rate of the nearby user with ipSIC converges to a throughput ceiling in the high SNR region. With the increasing of RIS elements, the RIS-TW-NOMA networks are capable of providing enhanced ergodic rates.
  \item
   We further discuss the system throughput and energy efficiency of RIS-TW-NOMA networks in both delay-limited and delay-tolerant transmission modes. It is shown that the system throughput of RIS-TW-NOMA is superior to that of RIS-TW-OMA. We also confirm that the energy efficiency of RIS-TW-NOMA with pSIC is superior to that with ipSIC in the high SNR region. In addition, the energy efficiency of RIS-TW-NOMA and
   RIS-TW-OMA networks outperform that of TWR-OMA networks.
\end{enumerate}

\subsection{Organization}
The rest of this paper is organized as follows. In Section II, the system model for RIS-TW-NOMA is introduced in detail.
The exact expressions of outage probability is derived and the system throughput is studied in Section III .
Fruthermore, Section IV derives the exact expressions of ergodic rate and obtains the high SNR slopes.
Section V investigates the energy efficiency for RIS-TW-NOMA networks.
 Numerical results are presented and verified in Section VI, which is followed by the conclusions in Section VII.
\section{System Model}\label{System Model}
\begin{figure}[t!]             
\includegraphics[width=3.3 in, height=2.1in]{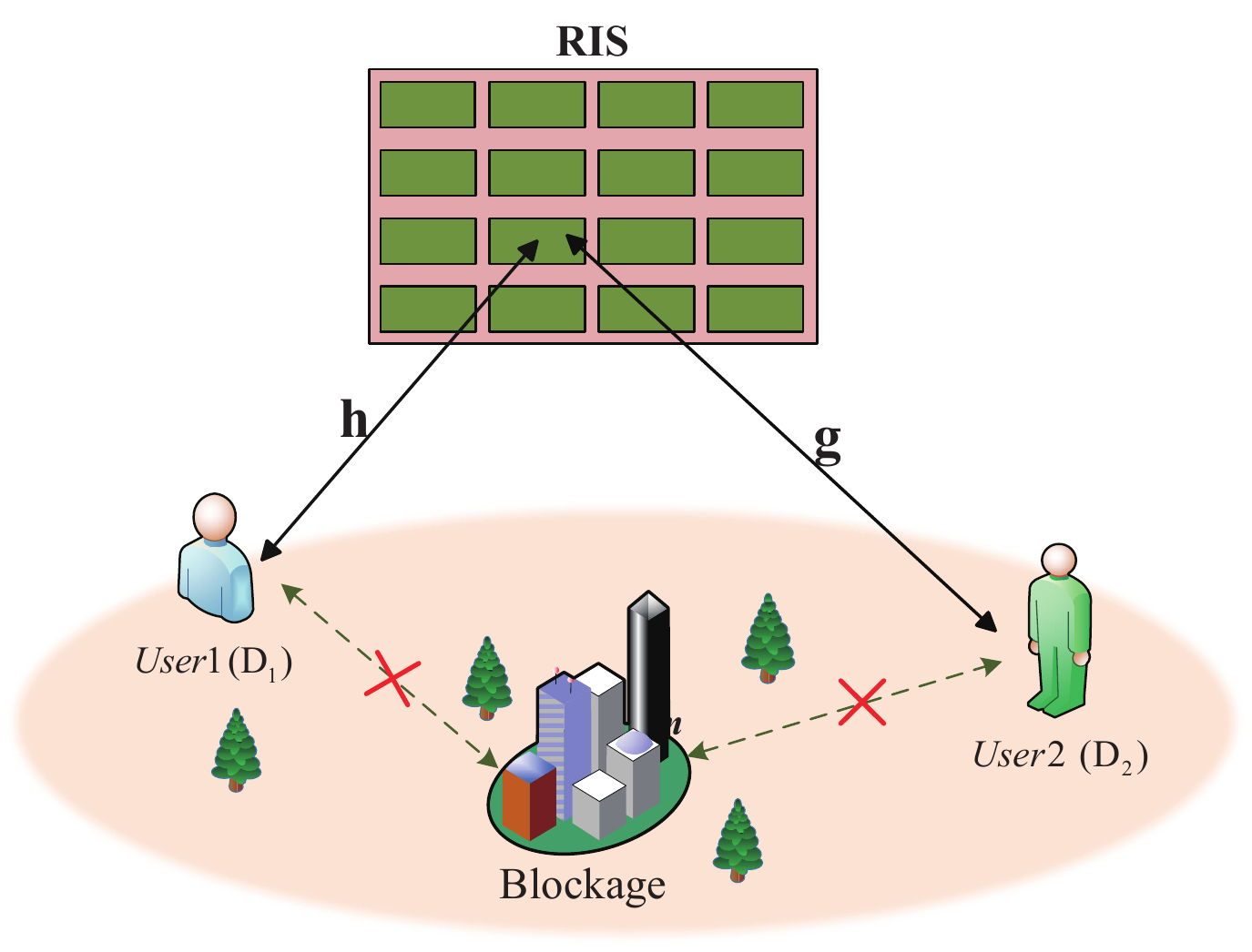}
 \caption{\small The system model of RIS-TW-NOMA networks, where a pair of blocked
users can exchange information by RIS.}
\label{Fig. 1}
\end{figure}

We consider a RIS-TW-NOMA system as illustrated in Fig. 1, where a pair of users, i.e., the nearby user ${D_1}$ and distant user ${D_2}$, are able to exchange their information assisted by RIS\footnote{It is noteworthy that the system model considered can also extended to the scenario of multiple users, i.e.., multiple pairs of users exchange their information via a RIS. each paired user is regarded as a group and within each group, the superposition coding and SIC are employed to decode the desired signals. Across the groups, the OMA scheme is employed and there are no co-channel interference among different pairs of users.}  Two users transmit their signals to the RIS, and then the superposed signals are reflected to ${D_1}$ and ${D_2}$ by the RIS, simultaneously.
We assume that the direct link between ${D_1}$ and ${D_2}$ is assumed strongly attenuated, and the communication can only be established by RIS.
The users ${D_1}$ and ${D_2}$ are individually equipped
with two antennas (a receiving antenna and a transmitting antenna) to enable a full-duplex mode. We consider that the RIS consists of $M$ elements.
The wireless communication links for the RIS-TW-NOMA networks are modeled as Rayleigh fading channels and interfered by additive white Gaussian noise (AWGN).The vectors of channel coefficients between the two users and RIS are given as ${\bf{h}} = {\left[ {{h_1},...,{h_m},...,{h_M}} \right]^H}$\
and ${\bf{g}}={\left[ {{g_1},...,{g_m},...,{g_M}} \right]^H}$, where ${{{h}}_m } \sim {\cal C}{\cal N}\left( {0,1} \right)$ and ${{{g}}_m } \sim {\cal C}{\cal N}\left( {0,1} \right)$ denote complex channel coefficients between the two users and the $m$-th RIS element.
The effective cascade channel gains can be written as ${{{\bf{h}}^{H}}{\bf {\Phi}}  {\bf{g}}}$, where ${\bf {\Phi}}  {\rm{ = diag}}\left( {\beta {e^{j{\theta _1}}},...,\beta {e^{j{\theta _m}}},...,\beta {e^{j{\theta _M}}}} \right)$ denotes the $M \times M$ diagonal phase shifting matrix with its $M$ main diagonal elements representing the RIS elements.

Referring to \cite{Ding2016relays,Yue8302918}, two NOMA users are categorized by their quality of service (QoS) requirements. The distant user ${D_2}$ is to be served for small packet transmission with a low data rate, should be given higher priority, while the nearby user ${D_1}$ is to be served opportunistically with a high data rate.
In particular, to ensure the fairness between ${D_1}$  and ${D_2}$,
the corresponding power allocation coefficient is represented by  ${a_i}$, which satisfies the relationship ${a_1} + {a_2} = 1$ and ${a_2} > {a_1}$ since ${D_2}'s$ QoS requirements are given higher priority.
It is worth noting that we assume that all perfect channel state information at the users is available.

\subsection{RIS-TW-NOMA }
Assuming only first-order reflection from RIS, the reflected signal at time \emph{t} is ${y_R}\left( t \right) = \sqrt {{P_u}{a_2}} {\bf{g}}{x_2}(t) + \sqrt {{P_u}{a_1}} {\bf{h}}{x_1}(t)$.
At this time, each user receives a superposition of the two signals via the RIS. Thus, the signal received at ${D_1}$ reflected by RIS is given by
\begin{align}\label{express1 }
{y_1}(t) = &\,\,{{\bf{h}}^{H}}{\bf{{\bf {\Phi}}  }}\left[ {\sqrt {{P_u}{a_2}} {\bf{g}}{x_2}(t) + \sqrt {{P_u}{a_1}} {\bf{h}}{x_1}(t)} \right]\nonumber\\
& + {x_{I_1}}(t)+ {{{n}}_{D_1}}(t),
\end{align}
where ${P_u}$ denotes the normalized transmission power, ${x_{I_i}}(t)$ is the received residual self-interference generated by multiple cancellation stages with
distribution ${\cal C}{\cal N}\left( {0,\sigma _{{I_i}}^2} \right)$ , $i \in \left\{ {1,2} \right\}$.
 ${n_{D_1}}(t)$ is the AWGN at ${D_1}$ with the mean power $\sigma _{{I_1}}^2$.
Based on NOMA principle, the received signal-to-interference-plus-noise ratio
(SINR) at ${D_1}$ to decode the ${D_2}'s$ information ${x_2}$ can be written as
\begin{align}\label{express2 }
{\gamma _{{D_1} \to {x_2}}} = \frac{{{P_u}{a_2}{{\left| {{{\bf{h}}^{H}}{\bf {\Phi}}  {\bf{g}}} \right|}^2}}}{{\varepsilon {P_u}{{\left| {{g_h}} \right|}^2} + \sigma _{{I_1}}^2{\rm{ + }}\sigma _{{n_1}}^2}},
\end{align}
where $0 \le \varepsilon\le1 $. ${\varepsilon= 0}$\ and  $0 < \varepsilon\le1 $ denotes the situations of pSIC and ipSIC.
Without loss of generality, assuming that the residual interference from ipSIC is modeled as the Rayleigh fading and corresponding complex channel coefficient is denoted by ${{g_{h}} } \sim {\cal C}{\cal N}\left( {0,\sigma _{{g_h}}^2}\right)$.

The signal received at ${D_2}$ reflected by RIS is given by
\begin{align}\label{express4}
{y_2}(t) =&\,\,{{\bf{g}}^{H}}{\bf{{\bf {\Phi}}  }}\left[ {\sqrt {{P_u}{a_2}} {\bf{g}}{x_2}(t) + \sqrt {{P_u}{a_1}} {\bf{h}}{x_1}(t)} \right]\nonumber\\
 &+ {x_{{I_2}}}(t) + {n_{D_2}}(t),
\end{align}
where ${n_{D_2}}(t)$ is the AWGN at ${D_2}$ with the mean power $\sigma _{{I_2}}^2$. Based on the principle of NOMA, the received SINR at ${D_2}$
to detect the ${D_1}'s$ information ${x_1}$ can be written as
\begin{align}\label{express5}
{\gamma _{{D_2} \to {x_1}}} = \frac{{{P_u}{a_1}{{\left| {{{\bf{h}}^{H}}{\bf {\Phi}}  {\bf{g}}} \right|}^2}}}{{\sigma _{{I_2}}^2 + \sigma _{{n_2}}^2}}.
\end{align}

\subsection{RIS-TW-OMA }
In this subsection, the RIS-TW-OMA scheme is one of the benchmark schemes. The entire communication process involves two time slots.
In the first slot, ${D_2}$ sends the signal ${x_2}$ to ${D_1}$ through RIS, and the signal received at ${D_1}$ is given by
\begin{align}\label{express11}
y_{{D_1}}^1(t) = {{\bf{h}}^H}{\bf{\Phi g}}\sqrt {{P_u}{a_2}} {x_2}(t) + {n_{{D_1}}}(t).
\end{align}
For the RIS-TW-OMA case, the SINR at ${D_1}$ to decode the ${D_2}'s$ information ${x_2}$ can be expressed as
\begin{align}\label{express11}
\gamma _{{D_1},{x_2}}^{1} = \frac{{{P_u}{a_2}{{\left| {{{\bf{h}}^{H}}{\bf {\Phi}}  {\bf{g}}} \right|}^2}}}{{ \sigma _{{n_1}}^2}}.
\end{align}

In the second slot, ${D_1}$ sends the signal ${x_1}$ to ${D_2}$ through RIS, and the signal received at ${D_2}$ is given by
\begin{align}\label{express11}
y_{{D_2}}^1(t) = {{\bf{h}}^H}{\bf{\Phi g}}\sqrt {{P_u}{a_1}} {x_1}(t) + {n_{{D_2}}}(t).
\end{align}
Hence, the decoding SINR at ${D_2}$ can be expressed as
\begin{align}\label{express11}
\gamma _{{D_2},{x_1}}^{1} = \frac{{{P_u}{a_1}{{\left| {{{\bf{h}}^{H}}{\bf {\Phi}}  {\bf{g}}} \right|}^2}}}{{ \sigma _{{n_2}}^2}},
\end{align}

\subsection{TWR-OMA }
In this subsection, the TWR-OMA scheme is regarded as another benchmark for comparison. The TWR-OMA system consists of a pair of users and a DF relay. The signals for the two users are divided into two time slots.
One user transmits its signal to relay in the first slot, and then relay transmits the signal to another user in the second slot.
Hence, the signal received at ${D_1}$  can be given by
\begin{align}\label{express11}
y_{{D_1}}^2(t) = h\sqrt {{P_u}{a_2}} {x_2}(t) + {n_{{D_1}}}(t).
\end{align}
Based on the above expression, the SINR at ${D_1}$ to decode the ${D_2}'s$ signal ${x_2}$ is given by
\begin{align}\label{express12}
\gamma _{{D_1},{x_2}}^{2} = \frac{{{P_u}{a_2}{{\left| h \right|}^2}}}{{\sigma _{{n_1}}^2}}.
\end{align}
For the the distant user, the signal received can be given by
\begin{align}\label{express11}
y_{{D_2}}^2(t) = h\sqrt {{P_u}{a_1}} {x_1}(t) + {n_{{D_2}}}(t).
\end{align}
Therefore, the decoding SINR for ${D_2}$ is as follows
\begin{align}\label{express12}
\gamma _{{D_2},{x_1}}^{2} = \frac{{{P_u}{a_1}{{\left| h \right|}^2}}}{{\sigma _{{n_2}}^2}}.
\end{align}

\section{OUTAGE PROBABILITY}
In wireless communication networks, the theoretical analyses of outage probability are the crucial works, which can guide the design and performance optimization of wireless communication systems' practical. In the next part, the outage behaviors of $D_1$ and $D_2$ for RIS-TW-NOMA networks are investigated in details.

\subsection{Outage Probability for RIS-TW-NOMA}
Currently, two main types of phase-shifting designs, i.e., coherent phase shifting and random phase shifting, are taken into consideration [28]. More specifically, the coherent phase shifting has the ability to enhance the system performance, where the phase shifting of each reflecting element is matched with the phases of its incoming and outgoing fading channels. The application of coherent phase shifting to RIS-TW-NOMA networks is capable of  simplifying computational complexity and providing distinct analytical results. It is worth pointing out that the random phase-shifting of RIS can also affect the outage behaviors and ergodic rate, which will be set aside in our future works. In particular, the channel vector, i.e.,  ${\tilde \chi } = {{\bf{h}}^H}{\bf {\Phi}}  {\bf{g}}{\rm{ = }}\sum\limits_{m = 1}^M {\left| {{h_m}{g_m}{e^{ - j{\theta _m}}}} \right|}$ of RIS-TW-NOMA networks is transformed to $\chi  = \sum\limits_{m = 1}^M {\left| {{h_m}{g_m}} \right|}$.
On the basis of ~\cite[Eq. (7)]{2014Outage}, the PDF of $\left| {{h_m}{g_m}} \right|$ can be expressed as \begin{align}\label{express15}
{f_{\left| {{h_m}{g_m}} \right|}}(x) = 4x{K_0}\left({2x} \right),
\end{align}
where ${K_0}\left(  \cdot  \right)$ is the modified Bessel function of the second kind with order zero.

The mean and variance of $\left| {{h_m}{g_m}} \right|$ can be formulated as
\begin{align}\label{express16}
{\mu _{\left| {{h_m}{g_m}} \right|}} = \int_0^\infty  {4{x^2}{K_0}\left( {2x} \right)} dx = \frac{\pi }{4},
\end{align}
and
\begin{align}\label{express17}
\sigma _{_{\left| {{h_m}{g_m}} \right|}}^2 = \int_0^\infty  {4{x^3}{K_0}\left( {2x} \right)} dx = 1 - \frac{{{\pi ^2}}}{{16}},
\end{align}
respectively. Since the random variables (RVs) $\left| {{h_m}{g_m}} \right|$ for all $m \in [1,M]$ are i.i.d., the $\chi$ can be given by
\begin{align}\label{express18}
\chi  = \left| {\sum\limits_{m = 1}^M {{h_m}{g_m}} } \right| \sim  {\cal N}( {M{\mu _{\left| {{h_m}{g_m}} \right|}},M\sigma _{\left| {{h_m}{g_m}} \right|}^2} ).
\end{align}
Noting that even though using the design of coherent phase-shifting,
it is challenge to derive the closed-form expression of the outage probability for $D_1$ and $D_2$ in RIS-TW-NOMA networks. Furthermore,
the CLT-based method is further employed to derive the approximate outage probability expressions.
In particular, the channel vector of RIS-TW-NOMA is $\chi  = {{\bf{h}}^H}{\bf {\Phi}}  {\bf{g}}{\rm{ = }}\sum\limits_{m = 1}^M {\left| {{h_m}{g_m}{e^{ - j{\theta _m}}}} \right|}$.
For the coherent phase shifting design, the phase shifts of the RIS are matched with the phases of the RIS fading gains, the $\chi$ can be further expressed as
$\sum\limits_{m = 1}^M {\left| {{h_m}{g_m}} \right|}$.
By using the CLT, the RV $\frac{{\chi  - M{\mu _{\left| {{h_m}{g_m}} \right|}}}}{{\sqrt M {\sigma _{\left| {{h_m}{g_m}} \right|}}}}$ obeys the standard normal distribution ${\cal N}\left( {0,1} \right)$.
When \emph{M} is sufficiently large, it can be further approximated as
a Gaussian random variable, which can be given by

\begin{align}\label{express19}
X = \sqrt M \left( {\frac{\chi }{M} - {\mu _{_{\left| {{h_m}{g_m}} \right|}}}} \right)\sim{\cal N}( {0,\sigma _{_{_{\left| {{h_m}{g_m}} \right|}}}^2}).
\end{align}

The outage behavior is an essential metric of performance analysis in RIS-TW-NOMA networks. The outage probability
of each user can be expressed as ${P_{{D_i}}} = {{\rm{P}}{\rm{r}}}[\gamma  < {\gamma _{th_i}}]$,
where ${\gamma _{t{h_i}}}$ is the SINR threshold expressed as ${\gamma _{t{h_i}}} = {2^{{R_i}}} - 1$ with ${R_i}$ being the target rate at the user to detect ${x_i}$.

\subsubsection{${D_1}$ of outage probability}\label{${D_1}$ of outage probability}
The SIC scheme is executed at ${D_1}$ by decoding and expurgating the ${D_2}'s$ information ${x_2}$
before it detects its own signal. According to the above explanation, the outage probability can be denoted by
\begin{align}\label{express20}
P_{{D_1}}^{ipSIC} = {{\rm{P}}{\rm{r}}}[{\gamma _{{D_1} \to {x_2}}} < {\gamma _{t{h_2}}}].
\end{align}

\begin{theorem} \label{theorem:1}By using the CLT, the exact expression of outage probability for ${D_1}$ with ipSIC in RIS-TW-NOMA networks is given by
\begin{align}\label{express21}
P_{{D_1}}^{ipSIC} = \frac{1}{2} + \frac{1}{{\sqrt \pi  }}\phi\left[ {\frac{{\sqrt M \left( {\frac{\tau }{M} - {\mu _{_{\left| {{h_m}{g_m}} \right|}}}} \right)}}{{\sqrt 2 {\sigma _{_{\left| {{h_m}{g_m}} \right|}}}}}} \right],
\end{align}
where $\phi \left( x \right) \!\!\buildrel \Delta \over =\!\! \int_0^x {{e^{ - {t^2}}}} dt $ and $\tau \!\! =\!\! \sqrt {\frac{{{\gamma _{t{h_2}}}}}{{{P_u}{a_2}}}(\varepsilon {P_u}\sigma _{{g_h}}^2 + \sigma _{{I_1}}^2{\rm{ + }}\sigma _{{n_1}}^2)} $.
\begin{proof}
See Appendix A.
\end{proof}
\end{theorem}

\begin{corollary} \label{corollary:1} For the particular case $\varepsilon =0$, we can derive the exact expression of outage probability for ${D_1}$ with pSIC in RIS-TW-NOMA networks as

\begin{align}\label{express23}
P_{{D_1}}^{pSIC} = \frac{1}{2} + \frac{1}{{\sqrt \pi  }}\phi \left( {\frac{{\sqrt M \left( {\frac{\psi }{M} - {\mu _{_{\left| {{h_m}{g_m}} \right|}}}} \right)}}{{\sqrt 2 {\sigma _{_{\left| {{h_m}{g_m}} \right|}}}}}} \right),
\end{align}
where $\psi  = \sqrt {\frac{{{\gamma _{t{h_2}}}}}{{{P_u}{a_2}}}(\sigma _{{I_1}}^2{\rm{ + }}\sigma _{{n_1}}^2)} $.
\end{corollary}

\subsubsection{${D_2}$ of outage probability}\label{${D_2}$ of outage probability}
 The outage event occurs when ${D_2}$ can delete the signal ${x_2}$ and does not decode the signal ${x_1}$ successfully, the outage probability of ${D_2}$ can be written as
\begin{align}\label{express24}
{P_{{D_2}}} = {{\rm{P}}{\rm{r}}}[{\gamma _{{D_2} \to {x_1}}} < {\gamma _{t{h_1}}}].
\end{align}

\begin{corollary} \label{corollary:3}The exact expression of outage probability for ${D_2}$ can be derived as
\begin{align}\label{express25}
{P_{{D_2}}} = \frac{1}{2} + \frac{1}{{\sqrt \pi  }}\phi \left[ {\sqrt {\frac{M}{{2(1 - \frac{{{\pi ^2}}}{{16}})}}} \left( {\frac{\beta }{M} - \frac{\pi }{4}} \right)} \right],
\end{align}
where $\beta  = \sqrt {\frac{{{\gamma _{t{h_1}}}}}{{{P_u}{a_1}}}(\sigma _{{I_2}}^2{\rm{ + }}\sigma _{{n_2}}^2)} $.
\end{corollary}

\subsection{Outage Probability for RIS-TW-OMA}
\subsubsection{${D_i}$ of outage probability}\label{${D_i}$ of outage probability}
For RIS-TW-OMA, the outage of ${D_i}$ is defined as the probability that the instantaneous SNR falls below a threshold SNR $\gamma _{t{h_l}}^i$.
Hence, the outage probability of ${D_i}$ in RIS-TW-OMA networks is denoted by
\begin{align}\label{express27}
P_{{D_i}}^i = {{\rm{P}}{\rm{r}}}[\gamma _{Di,{x_l}}^i < \gamma _{t{h_l}}^i],
\end{align}
where $\left( {i,l} \right) \in \left\{ {\left( {1,2} \right),\left( {2,1} \right)} \right\}$.
\begin{corollary} \label{corollary:4}Base on (23), we can derive the exact expression of outage probability for ${D_i}$  in RIS-TW-OMA networks as
\begin{align}\label{express28}
P_{{D_i}}^{i}
&= {{\rm{P}}{\rm{r}}}\left[ {X < \sqrt M \left( {\frac{{{\lambda _i}}}{M} - {\mu _{\left| {{h_m}{g_m}} \right|}}} \right)} \right]\nonumber\\
&= \frac{1}{2} + \frac{1}{{\sqrt \pi  }}\phi \left( {\sqrt {\frac{M}{{2(1 - \frac{{{\pi ^2}}}{{16}})}}} \left( {\frac{{{\lambda _i}}}{M} - \frac{\pi }{4}} \right)} \right),
\end{align}
where ${\lambda _i}{\rm{ = }}\sqrt {\frac{{\gamma _{t{h_l}}^{i}\sigma _{{n_i}}^2}}{{{P_u}{a_l}}}}$ and $\gamma _{t{h_l}}^{i}{\rm{ = }}{2^{{R_l}}} - 1$.
\end{corollary}

\subsection{Diversity Analysis}
In order to get more insights, we select the diversity order to evaluate the outage behaviors in the high SINR region \cite{Liu2017physicallayer}, which is mathematically described as
\begin{align}\label{express28}
d =  - \mathop {\lim }\limits_{\rho  \to \infty } \frac{{\log \left( {{P_\infty }(\rho )} \right)}}{{\log \rho }},
\end{align}
where $\rho $ denotes the transmit SNR. It is worth pointing out that ${P_u}$ stands for $\rho $ in this article.
It is not tractable to obtain diversity order through the exact expression of the outage probability. The CLT-based approximation method is not accurate in the high SNR region. Thus, we evaluate the performance boundaries for the outage probability. The upper bound on the outage probability is derived using the Bessel function's upper bound. In the high SNR region, the upper bound of the outage probability converges to zero. Therefore, the outage performance of the network at high SNR is accurately described by the upper bound of the outage probability.
\begin{theorem} \label{theorem:2}To approximate analysis, the upper bound for outage probability of ${D_1}$
with ipSIC in RIS-TW-NOMA networks can be given by
\begin{align}\label{express33}
P_{{D_1}}^{ipSIC} \le \frac{{{2^{ - L}}{\pi ^{\frac{M}{2}}}{\Gamma ^M}\left( {\frac{3}{2}} \right)}}{{(3L - 1)!}}\gamma \left( {3L,2\tau } \right),
\end{align}
where $M$ is an even number $L = \frac{M}{2}$ and $\gamma (s,x) = \int_0^x {{t^{s - 1}}} {{\rm{e}}^{ - t}}dt$ is the incomplete Gamma function.
\begin{proof}
See Appendix B.
\end{proof}
\end{theorem}
\begin{corollary} \label{corollary:2}Similar to (26),
the outage probability of ${D_1}$ with pSIC in RIS-TW-NOMA networks can be upper bounded by
\begin{align}\label{express34}
P_{{D_1}}^{pSIC} \le \frac{{{2^{ - L}}{\pi ^{\frac{M}{2}}}{\Gamma ^M}\left( {\frac{3}{2}} \right)}}{{(3L - 1)!}}\gamma \left( {3L,2\psi  } \right).
\end{align}
\end{corollary}
\begin{corollary} \label{corollary:4}The upper bound for outage probability of ${D_2}$
 in RIS-TW-NOMA networks can be represented by
\begin{align}\label{express34}
{P_{{D_2}}} \le \frac{{{2^{ - L}}{\pi ^{\frac{M}{2}}}{\Gamma ^M}\left( {\frac{3}{2}} \right)}}{{(3L - 1)!}}\gamma \left( {3L,2\beta } \right).
\end{align}
\end{corollary}
\begin{corollary} \label{corollary:7}
When ${P_u}$ tends to infinity, $\tau $ tends to $\sqrt {\frac{{{\gamma _{t{h_2}}}\varepsilon \sigma _{{g_h}}^2}}{{{a_2}}}}$,
the approximated expression for outage probability of ${D_1}$ with ipSIC in RIS-TW-NOMA networks is given by
\begin{align}\label{express31}
P_{{D_1}}^{\infty ,{\rm{i}}pSIC}& \approx {2^{ - L}}{\pi ^{\frac{M}{2}}}{\left[ {\Gamma \left( {\frac{3}{2}} \right)} \right]^M}\frac{{{2^{3L}}{\tau ^{3L}}}}{{(3L)!}}\nonumber\\
&= {2^{ - L}}{\pi ^{\frac{M}{2}}}{\left[ {\Gamma \left( {\frac{3}{2}} \right)} \right]^M}\frac{{{2^{3L}}}}{{(3L)!}}\sqrt {\frac{{{\gamma _{t{h_2}}}\varepsilon \sigma _{{g_h}}^2}}{{{a_2}}}}.
\end{align}
\end{corollary}
\begin{remark} \label{remark:1}
When ${P_u} \to \infty$, upon substituting (29) into (25), i.e., $d_{{D_1}}^{ipSIC}\!\!\!=\!\!-\!\!\mathop{\lim }\limits_{{P_u}\!\to \infty } \!\!\frac{{\log P_{{D_1}}^{\infty ,{\rm{i}}pSIC}}}{{\log {P_u}}}\!\!=\!\!0$,  we can obtain a zero diversity order of ${D_1}$ with ipSIC for RIS-TW-NOMA.
 This is due to the fact that the residual interference from ipSIC leads diversity order to zero.
\end{remark}

\begin{corollary} \label{corollary:8}
When $\psi$ tends to zero,
the approximated expression for outage probability of ${D_1}$ with pSIC is given by
\begin{align}\label{express34}
P_{{D_1}}^{\infty ,pSIC}\approx  {2^{ - L}}{\pi ^{\frac{M}{2}}}{\left[ {\Gamma \left( {\frac{3}{2}} \right)} \right]^M}\frac{{{2^{3L}}{\psi ^{3L}}}}{{(3L)!}} \propto \frac{1}{{{P_u}^{\frac{3}{4}M}}}.
\end{align}
\end{corollary}

\begin{remark} \label{remark:2}
Under pSIC conditions, upon substituting (30) into (25),  i.e., $d_{{D_1}}^{pSIC}\!\!\!\!=\!\!- \!\!\mathop {\lim }\limits_{{P_u} \to \infty } \!\!\frac{{\log {P_u}^{ - \frac{3}{4}M}}}{{\log {P_u}}}\!\!=\!\!\frac{3}{4}M$, the diversity order of ${D_1}$ with pSIC is equal to $\frac{3}{4}M$ for RIS-TW-NOMA.
One can observe that the diversity order of ${D_1}$ with pSIC is in connection with the number of RIS elements.
\end{remark}

\begin{corollary} \label{corollary:9}
In the high SNR region, $\beta $ tends to zero,
the approximated expression for outage probability of ${D_2}$ in RIS-TW-NOMA networks is given by
\begin{align}\label{express34}
P_{{D_2}}^\infty  \approx {2^{ - L}}{\pi ^{\frac{M}{2}}}{\left[ {\Gamma \left( {\frac{3}{2}} \right)} \right]^M}\frac{{{2^{3L}}{\beta ^{3L}}}}{{(3L)!}} \propto \frac{1}{{{P_u}^{\frac{3}{4}M}}}.
\end{align}
\end{corollary}

\begin{remark} \label{remark:2}
Similar to the proof of the diversity order of ${D_1}$ , ${d_{{D_2}}}\!\!=\!\!-\!\!\mathop {\lim }\limits_{{P_u} \to \infty }\!\!\frac{{\log P_{{D_2}}^\infty }}{{\log {P_u}}}\!\!=\!\!\frac{3}{4}M$, the diversity order of ${D_2}$ is equal to $\frac{3}{4}M$.
As can be observed that the diversity order of ${D_2}$ for RIS-TW-NOMA is equal to the diversity order of ${D_1}$ with pSIC for RIS-TW-NOMA.
\end{remark}

\subsection{Throughput Analysis}
In delay-limited transmission scenario, by using the derived results of outage probability above, the throughput expression for RIS-TW-NOMA scheme is obtained as below.
\begin{align}\label{express30}
R^\delta  = \left( {1 - P_{{D_1}}^\delta } \right){R_1} + \left( {1 - {P_{{D_2}}}} \right){R_2},
\end{align}
where ${\delta  \in \left( {{\rm{i}}pSIC,pSIC} \right)}$, $P_{{D_1}}^{pSIC}$, $P_{{D_1}}^{ipSIC}$ and $P_{{D_2}}$ can be obtained from (19), (20) and (22), respectively.

\section{ERGODIC RATE}
The ergodic rate is another critical metric for evaluating system performance.
This section will focuses our attention on analyzing the ergodic rate of users in RIS-TW-NOMA networks and RIS-TW-OMA networks under ipSIC/pSIC conditions seriously.

\subsection{Ergodic Rate for RIS-TW-NOMA}

\subsubsection{${D_1}$ of ergodic rate }\label{${D_1}$ of ergodic rate}
 Under the situation that ${D_1}$ detects ${x_2}$ successfully,
 the achievable rate of ${D_1}$ can be written as ${R_{_{{D_1}}}} = \log \left( {1 + {\gamma _{{D_1} \to {x_2}}}} \right)$.
Therefore, the ergodic rate of ${D_1}$ with ipSIC in RIS-TW-NOMA networks can be expressed as
\begin{align}\label{express30}
R_{{D_1}}^{ipSIC} =  \mathbb{E} \left[ {\log \left( {1 + {\frac{{{P_u}{a_2}{{\left| {{{\bf{h}}^{H}}{\bf {\Phi}}  {\bf{g}}} \right|}^2}}}{{\varepsilon {P_u}{{\left| {{g_h}} \right|}^2} + \sigma _{{I_1}}^2{\rm{ + }}\sigma _{{n_1}}^2}}}} \right)} \right],
\end{align}
where ${\varepsilon= 1}$. It is not easy to obtain a closed-form expression by deriving the above formula. Only an exact expression can be calculated and evaluated numerically by Matlab software.

\begin{theorem} \label{theorem:3}The exact expression for ergodic rate of ${D_1}$ with ipSIC in RIS-TW-NOMA networks can be derived as
\begin{align}\label{1}
\begin{split}
R_{{D_1}}^{ipSIC} = & \frac{1}{{\ln 2}}\int_0^\infty  {\frac{1}{{1 + y}}} \left\{ {\frac{1}{2} - \frac{1}{{\sqrt \pi  }}\phi \left[ {\sqrt {\frac{M}{{2(1 - \frac{{{\pi ^2}}}{{16}})}}} } \right.} \right.\\
&\left. {\left. { \times \left( {\sqrt {\frac{{y(\varepsilon {P_u}\sigma _{{g_h}}^2 + \sigma _{{I_1}}^2{\rm{ + }}\sigma _{{n_1}}^2)}}{{P{a_2}}}} \frac{1}{M} - \frac{\pi }{4}} \right)} \right]} \right\}dy.
  \end{split}
\end{align}
\begin{proof}
See Appendix C.
\end{proof}
\end{theorem}

\begin{corollary} \label{corollary:9} For the particular case with substituting $\varepsilon =0$ into (33), the exact expression for ergodic rate of ${D_1}$ with pSIC in RIS-TW-NOMA networks is given by
\begin{align}\label{1}
\begin{split}
R_{{D_1}}^{pSIC}
=& \frac{1}{{\ln 2}}\int_0^\infty  {\frac{1}{{1 + y}}} \left\{ {\frac{1}{2} - \frac{1}{{\sqrt \pi  }}\phi \left[ {\sqrt {\frac{M}{{2(1 - \frac{{{\pi ^2}}}{{16}})}}} } \right.} \right.\\
&\left. {\left. { \times \left( {\sqrt {\frac{{(\sigma _{{I_1}}^2{\rm{ + }}\sigma _{{n_1}}^2)y}}{{{P_u}{a_2}}}} \frac{1}{M} - \frac{\pi }{4}} \right)} \right]} \right\}dy.
\end{split}
\end{align}
\end{corollary}

\subsubsection{${D_2}$ of ergodic rate }\label{${D_2}$ of ergodic rate}
The achievable rate of ${D_2}$ can be written as ${R_{_{{D_2}}}} = \log \left( {1 + {\gamma _{{D_2} \to {x_1}}}} \right)$. Based on (4),
the ergodic rate of ${D_2}$ in RIS-TW-NOMA networks is further can be expressed as
\begin{align}\label{express30}
{R_{{D_2}}} =  \mathbb{E} \left[ {\log \left( {1 +{\frac{{{P_u}{a_1}{{\left| {{{\bf{h}}^{H}}{\bf {\Phi}}  {\bf{g}}} \right|}^2}}}{{\sigma _{{I_2}}^2 + \sigma _{{n_2}}^2}}}} \right)} \right].
\end{align}

\begin{corollary} \label{corollary:4}By virtue of (36), the exact expression for ergodic rate of ${D_2}$ in RIS-TW-NOMA networks is given by

\begin{align}\label{1}
\begin{split}
{R_{{D_2}}}
=& \frac{1}{{\ln 2}}\int_0^\infty  {\frac{1}{{1 + y}}} \left\{ {\frac{1}{2} - \frac{1}{{\sqrt \pi  }}\phi \left[ {\sqrt {\frac{M}{{2(1 - \frac{{{\pi ^2}}}{{16}})}}} } \right.} \right.\\
&\left. {\left. { \times \left( {\sqrt {\frac{{(\sigma _{{I_2}}^2{\rm{ + }}\sigma _{{n_2}}^2)y}}{{{P_u}{a_1}}}} \frac{1}{M} - \frac{\pi }{4}} \right)} \right]} \right\}dy.
\end{split}
\end{align}
\end{corollary}

\subsection{Ergodic Rate for RIS-TW-OMA}
For RIS-TW-OMA networks, the ergodic rate of ${D_i}$ can be expressed as
\begin{align}\label{express30}
R_{{D_i}}^i =  \mathbb{E} \left[ {\frac{1}{2}\log \left( {1 + {\frac{{{P_u}{a_l}{{\left| {{{\bf{h}}^{H}}{\bf {\Phi}}  {\bf{g}}} \right|}^2}}}{{\sigma _{{n_i}}^2}}}} \right)} \right].
\end{align}
Similar to the proof of Theorem 3, the exact expression for ergodic rate of ${D_i}$ in RIS-TW-OMA networks can be attained.
\begin{corollary} \label{corollary:5}The exact expression for ergodic rate of ${D_i}$ in RIS-TW-OMA networks is given by
\begin{align}\label{1}
\begin{split}
R_{{D_i}}^i
=& \frac{1}{{2\ln 2}}\int_0^\infty  {\frac{1}{{1 + y}}} \left\{ {\frac{1}{2} - \frac{1}{{\sqrt \pi  }}\phi \left[ {\sqrt {\frac{M}{{2(1 - \frac{{{\pi ^2}}}{{16}})}}} } \right.} \right.\\
 &\times \left. {\left. {\left( {\sqrt {\frac{{{\rm{y}}\sigma _{{n_i}}^2}}{{{P_u}{a_l}}}} \frac{1}{M} - \frac{\pi }{4}} \right)} \right]} \right\}dy.
\end{split}
\end{align}

\end{corollary}
\subsection{Slope Analysis}
In order to gather deep insights for the communication networks performance, the high SNR slope is the critical parameter to evaluate the ergodic rate in the high SNR region,
which is able to describe how the ergodic rate
changes with the independent variable. The expression of high SNR slope can be written as
\begin{align}\label{express28}
S = \mathop {\lim }\limits_{\rho  \to \infty } \frac{{{R_\infty }(\rho )}}{{\log (\rho )}},
\end{align}
where ${{R_\infty }(\rho )}$ denotes the asymptotic ergodic rate in the high SNR region,
$\rho $ denotes the transmit SNR, it is worth pointing out that ${P_u}$ stands for $\rho $ in this article.

For the particular case with $\varepsilon =0$, the exact expression for ergodic rate in the high SNR region is not easy to obtain the approximation.
We use the upper bound of the ergodic rate to find its slope. When $\log \left( {1 + {x^2}} \right)$ is a concave function, a simple upper bound derived from Jensen's inequality is provided as
\begin{align}\label{express31}
R_{{D_1}}^{pSIC} \le \log \left[ {1 +  \mathbb{E}\left( {\frac{{{P_u}{a_2}{{\left| {{{\bf{h}}^{H}}{\bf {\Phi}}  {\bf{g}}} \right|}^2}}}{{\sigma _{{I_1}}^2 + \sigma _{{n_1}}^2}}} \right)} \right].
\end{align}
With the aid of (14) and (15), the upper bound ergodic rate of ${D_1}$ with pSIC in RIS-TW-NOMA networks can be expressed as
\begin{align}\label{express31}
R_{{D_1}}^{pSIC,up}\! =\! \log \left[ {1\! +\! \left( {\frac{{{P_u}{a_2}\left[ {{{\left( {\pi M} \right)}^2} + 16M - M{\pi ^2}} \right]}}{{16\left( {\sigma _{{I_1}}^2 + \sigma _{{n_1}}^2} \right)}}} \right)} \right].
\end{align}

\begin{remark} \label{remark:5}
Upon substituting (42) into (40) and using the L'Hospital's rule, the high SNR slope of ${D_1}$ with pSIC in RIS-TW-NOMA networks is equal to one.
\end{remark}

\begin{remark} \label{remark:5}
Similar to (42), by using Jensen's inequality, the upper bound ergodic rate of ${D_2}$ in RIS-TW-NOMA networks can be obtained.
 With the aid of (40) and using the L'Hospital's rule, the high SNR slope of ${D_2}$ in RIS-TW-NOMA networks is equal to one.
\end{remark}

\subsection{Throughput Analysis}
According to the derived results of outage probability above, the throughput expression in delay-tolerant transmission
 for RIS-TW-NOMA are obtained as below
\begin{align}\label{express30}
{R_\xi } = R_{{D_1}}^\delta  + {R_{{D_2}}},
\end{align}
where ${\delta  \in \left( {{\rm{i}}pSIC,pSIC} \right)}$, $R_{{D_1}}^{pSIC}$, $R_{{D_1}}^{ipSIC}$ and $R_{{D_2}}$ can be obtained from (34), (35) and (37), respectively.

\section{ENERGY EFFICIENCY}
The energy efficiency is a salient performance metric related to the ergodic rate and throughput.
 Based on the above analysis, the energy efficiency of RIS-TW-NOMA networks is characterized in this section.
We consider the energy efficiency performance for TWR-OMA and RIS-TW-OMA schemes as the benchmarks for comparison. The definition of energy efficiency can be given by

\begin{align}\label{28}
{\rm{Energy}}\;{\rm{efficiency = }}\frac{{{\rm{Total}}\;{\rm{data}}{\mkern 1mu} {\rm{rate}}}}{{{\rm{Total}}\;{\rm{energy}}{\mkern 1mu} {\kern 1pt} {\rm{consumption}}}}{\rm{.}}
\end{align}

 In RIS-TW-NOMA networks, the total power consumption is composed of the hardware static power dissipated at the RIS and user terminals \cite{8741198,Yue2021NOMARIS},
 and the energy efficiency of RIS-TW-NOMA is interpreted as the sum data rate divided by the total power consumption and is expressed as
\begin{align}\label{28}
{\eta _{EE}} = \frac{{R_{total}}}{{\varepsilon {P_u} + {P_{RIS}}{\rm{ + }}{P_1} + {P_2}}},
\end{align}
where ${R_{total}} \in ({R^\delta },{R_\xi })$, ${R^\delta }$ is the total data transmission
 rate in delay-limited transmission scenario, ${R_\xi }$ is the total data transmission
 rate in delay-tolerant transmission scenario.
The $\varepsilon  = {v^{ - 1}}$ and $v$ represents the efficiency of the transmit power amplifier,
${P_{RIS}} = K{P_k}\left( b \right)$ represents the static hardware loss power at the RIS, where ${P_k}\left( b \right)$ is the power consumption of each phase shifter having
\emph{b}-bit resolution \cite{7370753,8333733}. ${P_1}$ and ${P_2}$  are the static hardware loss power of ${D_1}$ and ${D_2}$, respectively.

\begin{table}[t]
\centering
\caption{Diversity order and high SNR slope for RIS-TW-NOMA and RIS-TW-OMA networks.}
\tabcolsep5pt
\renewcommand\arraystretch{1.85} 
\begin{tabular}{|c|c|c|c|c|}
\hline
\textbf{Mode } &  \textbf{SIC}  &  \textbf{User}  &  \textbf{D}  &  \textbf{S}  \\
\hline
\multirow{2}{*}{RIS-TW-OMA }  & \multirow{2}{*}{\raisebox{0.5mm}{------}} &   ${D_1}$  & $\frac{3}{4}M$ &$\frac{1}{2}$ \\
\cline{3-5}
                                &  &   ${D_2}$  & $\frac{3}{4}M$  & $\frac{1}{2}$ \\
\hline
\multirow{3}{*}{RIS-TW-NOMA } & ipSIC &   ${D_1}$  & 0 & - \\
\cline{2-5}
                                & pSIC &   ${D_1}$  & $\frac{3}{4}M$  & 1 \\
                                \cline{2-5}
                                & - &   ${D_2}$  & $\frac{3}{4}M$  & 1  \\
\hline
\end{tabular}
\label{parameter}
\end{table}

\section{NUMERICAL RESULTS}
In this section, the simulation results verify the rationality of the derived theoretical expressions for RIS-TW-NOMA networks.
Table II is the parameter of the Monte Carlo simulation, where BPCU denotes the short for a bit per channel use.
To verify the feasibility of the RIS-TW-NOMA networks, the outage probability, ergodic rate, energy efficiency, and system throughput are presented.
Without loss of generality, the power allocation coefficients of a pair of users are selected as ${a_1} = 0.2$ and ${a_2} = 0.8$, respectively.
 More specifically, we show the impact of target rate, residual interference, and RIS elements on the performance of RIS-TW-NOMA networks.
TWR-OMA and RIS-TW-OMA is shown as the comparison benchmarks.
 Furthermore, the performance of the three transmission schemes is evaluated through computer simulation.

\begin{table}[]
\centering \caption{ The parameters for numerical results}\label{runResult}
\begin{tabular}{|l|l|}
\hline
{\color[HTML]{333333} \textbf{Monte Carlo simulations repeated}} & ${10^6}$ iterations                                                                                          \\ \hline
Power allocation coefficients of NOMA                   & \begin{tabular}[c]{@{}l@{}}${a_1}$ = 0.2\\ ${a_2}$ = 0.8\end{tabular}                                   \\ \hline
Targeted data rates                                    & \begin{tabular}[c]{@{}l@{}}${R_1}$ = 2 BPCU\\ ${R_2}$ = 5 BPCU\\ ${R_3}$ = 6 BPCU\\ ${R_4}$ = 5.5 BPCU\end{tabular} \\ \hline
\end{tabular}
\end{table}
\subsection{Outage Probability}
Fig. \ref{Fig. 2} plots the outage probability of two users versus the transmit ${P_u}$ for TWR-OMA, RIS-TW-OMA, and RIS-TW-NOMA when $M= 8$. The exact fork and diamond curve for outage probability of ${D_1}$ with ipSIC/pSIC are plotted by (19) and (20), respectively.
The exact right triangle curve for outage probability of ${D_2}$ is plotted based on (22).
The Monte Carlo simulation outage probability curves are relatively identical to analytical results across the entire SNR range, proving our theoretical derivation's correctness.
One can observe that the outage probability with ipSIC converges to an error floor in the high SNR region and obtain a zero diversity order. This is due to the fact that the residual interference from ipSIC for RIS-TW-NOMA, which is also confirmed in \textbf{Remark 1}.
The asymptotic curves for outage probability of ${D_1}$ with ipSIC/pSIC and ${D_2}$ are drawn according to (26), (27), and (28), respectively. As shown from the figure, the CLT-based outage probability approximation curves are in accord with the exact outage probability curves in the low SNR region. The accuracy of the upper bounds for outage probability is higher than that of the CLT-based outage probability approximations in the high SNR region.
This is because that the skewed distribution of $\left| {{h_m}{g_m}} \right|$ causes errors in the two curves.
In the high SNR region, the CLT-based outage probability approximations are inaccurate.
Furthermore, the slopes of upper bounds for outage probability are the same as that of exact outage probability curves, which reveals that the upper bounds are accurate.
The exact outage probability cures of RIS-TW-OMA are plotted according to the analytical results in (24).
The critical observation is that the outage behaviors of RIS-TW-NOMA with pSIC are superior to that of TWR-OMA and RIS-TW-OMA, particularly in the high SNR region.
 The reasons is that the RIS-TW-NOMA networks can realize much better user fairness than TWR-OMA and RIS-TW-OMA networks for multiple users.
Compared with the ipSIC scheme, the RIS-TW-NOMA networks with pSIC can achieve better outage behavior.  As a result, it is important to consider the influence of ipSIC on the network performance for RIS-TW-NOMA in the practical scenario.

{\begin{figure}[t!]
\centering
\includegraphics[width=3.4in,  height=2.7in]{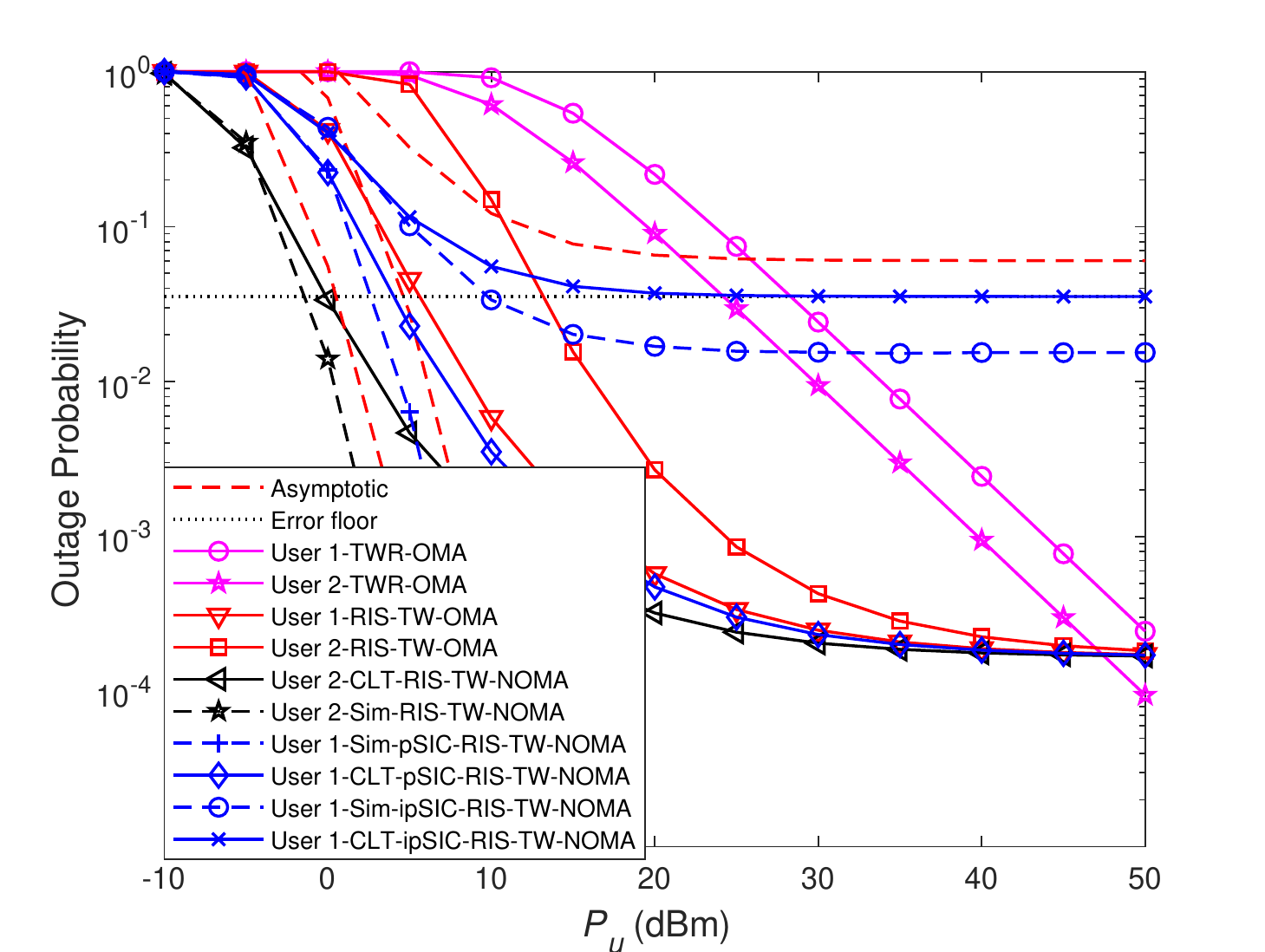}
\caption{Outage probability versus transmit ${P_u}$ for TWR-OMA, RIS-TW-OMA and RIS-TW-NOMA, with ${R_1} = 2$, ${R_2} = 5$ BPCU,
$\mathbb{E}\left\{ {{{\left| {{g_h}} \right|}^2}} \right\} =  - 6$ dB and
$\mathbb{E}\left\{ {{{\left| {{\sigma _{{I_i}}}} \right|}^2}} \right\} =  - 5$ dB.}
\label{Fig. 2}
\end{figure}

\begin{figure}[t!]
\centering
\includegraphics[width=3.4in,  height=2.7in]{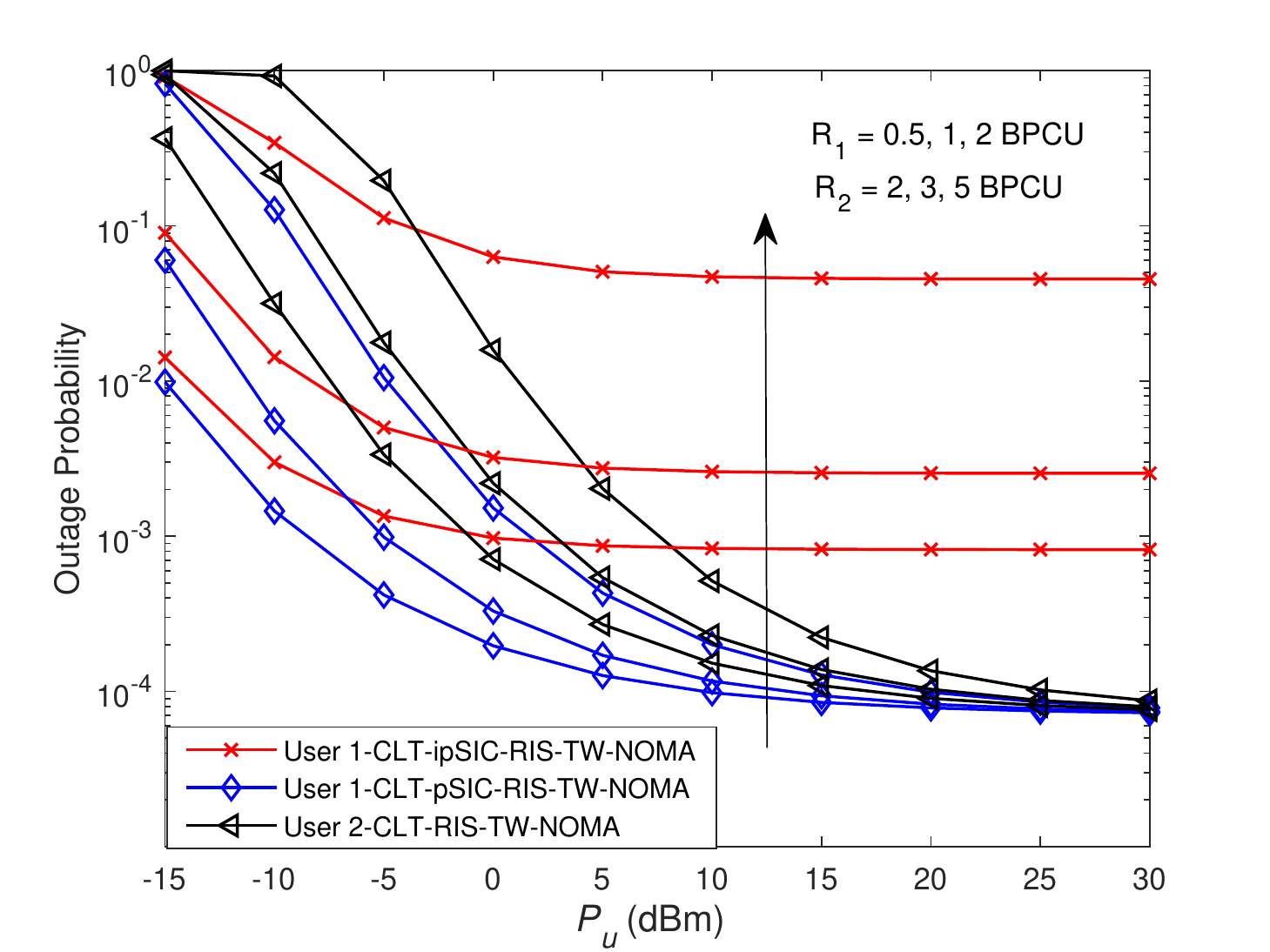}
\caption{Outage probability versus the transmit ${P_u}$, with the different target rate.}
\label{Fig. 4}
\end{figure}

\begin{figure}[t!]
\centering
\includegraphics[width=3.4in,  height=2.6in]{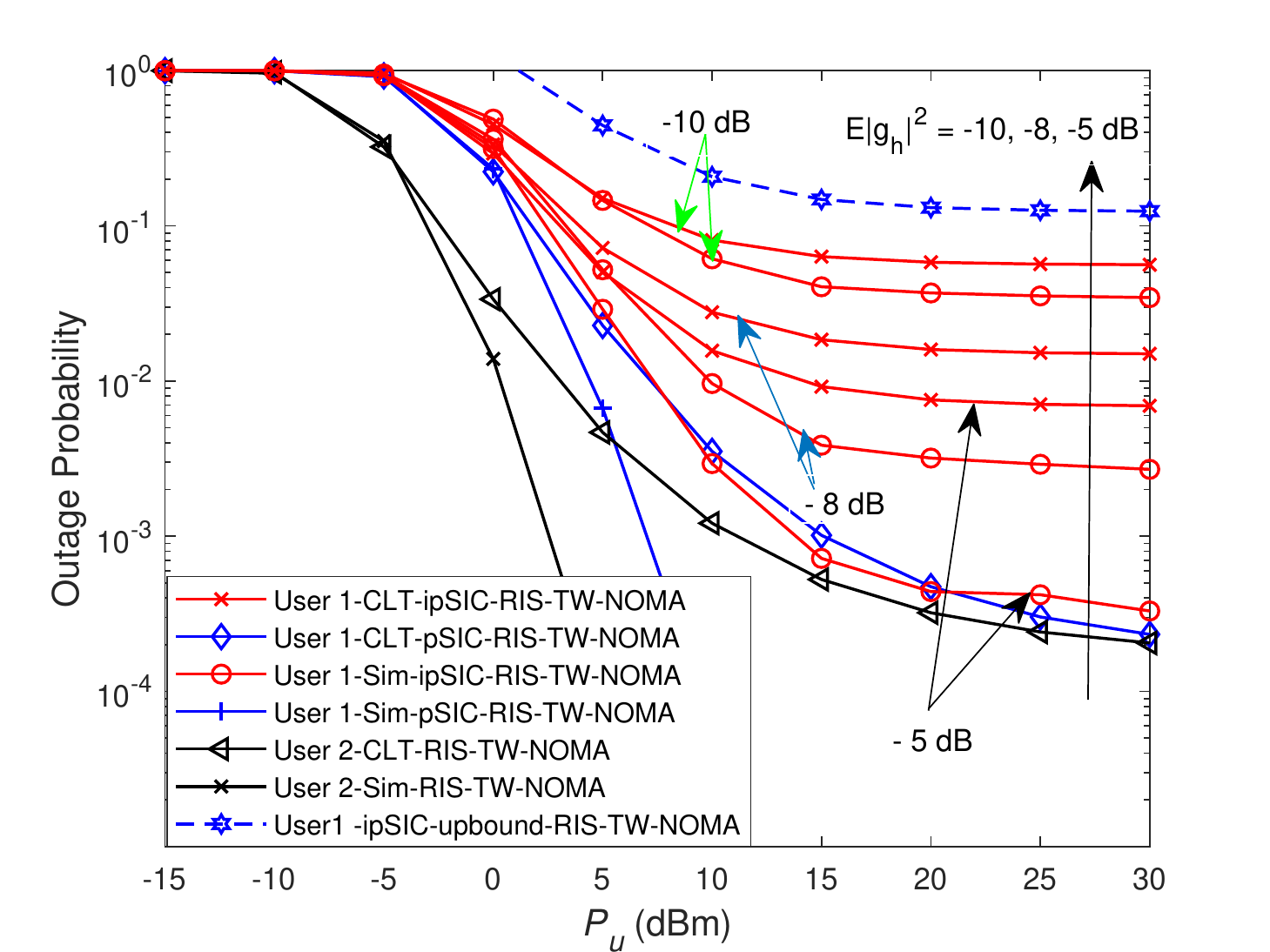}
\caption{Outage probability versus the transmit ${P_u}$, with the different residual interference.}
\label{Fig. 5}
\end{figure}

\begin{figure}[t!]
\centering
\includegraphics[width=3.3in,  height=2.6in]{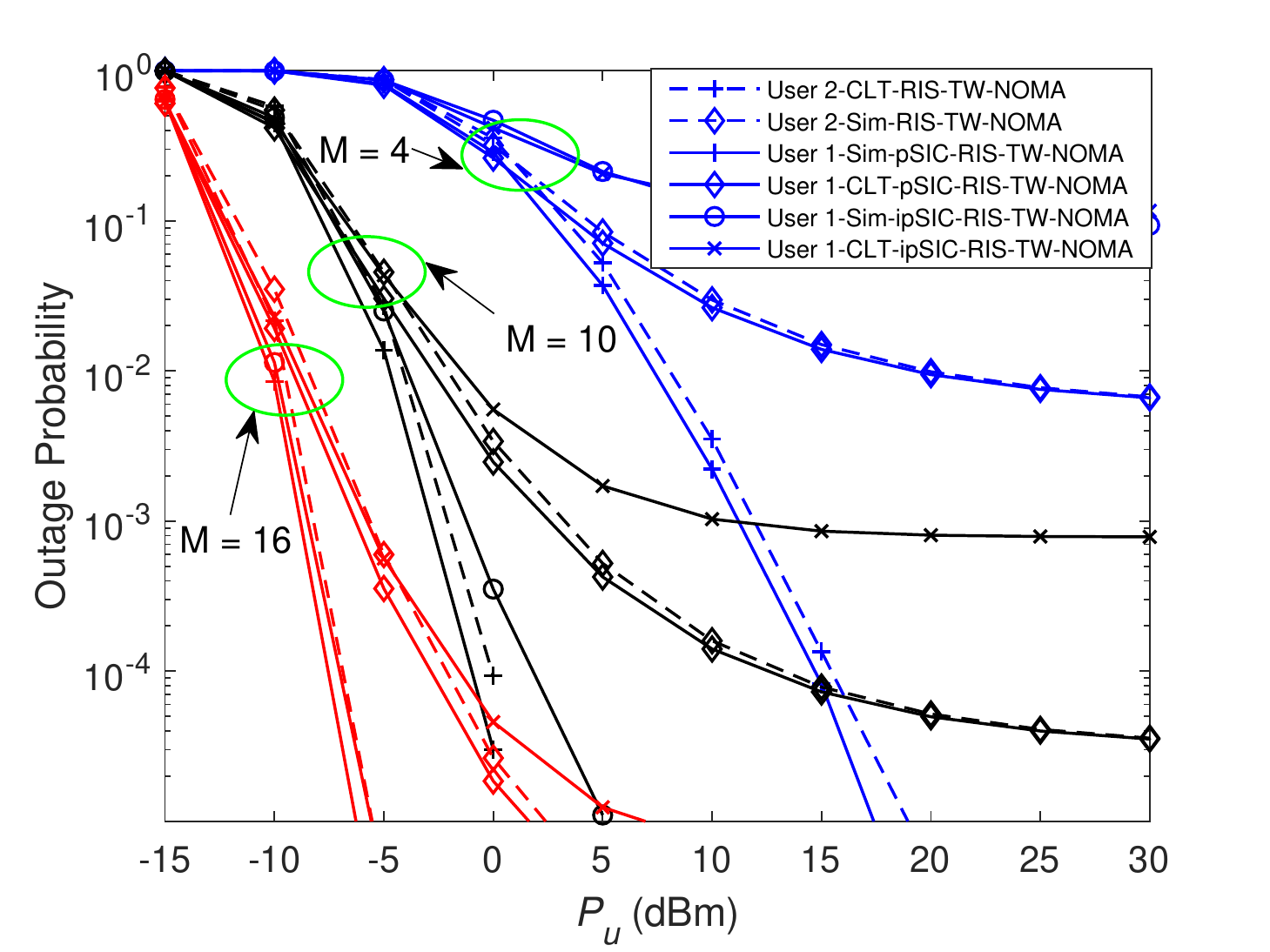}
\caption{Outage probability versus the transmit ${P_u}$, with the different \emph{M}.}
\label{Fig. 6}
\end{figure}

\begin{figure}[t!]
\centering
\includegraphics[width=3.3in,  height=2.6in]{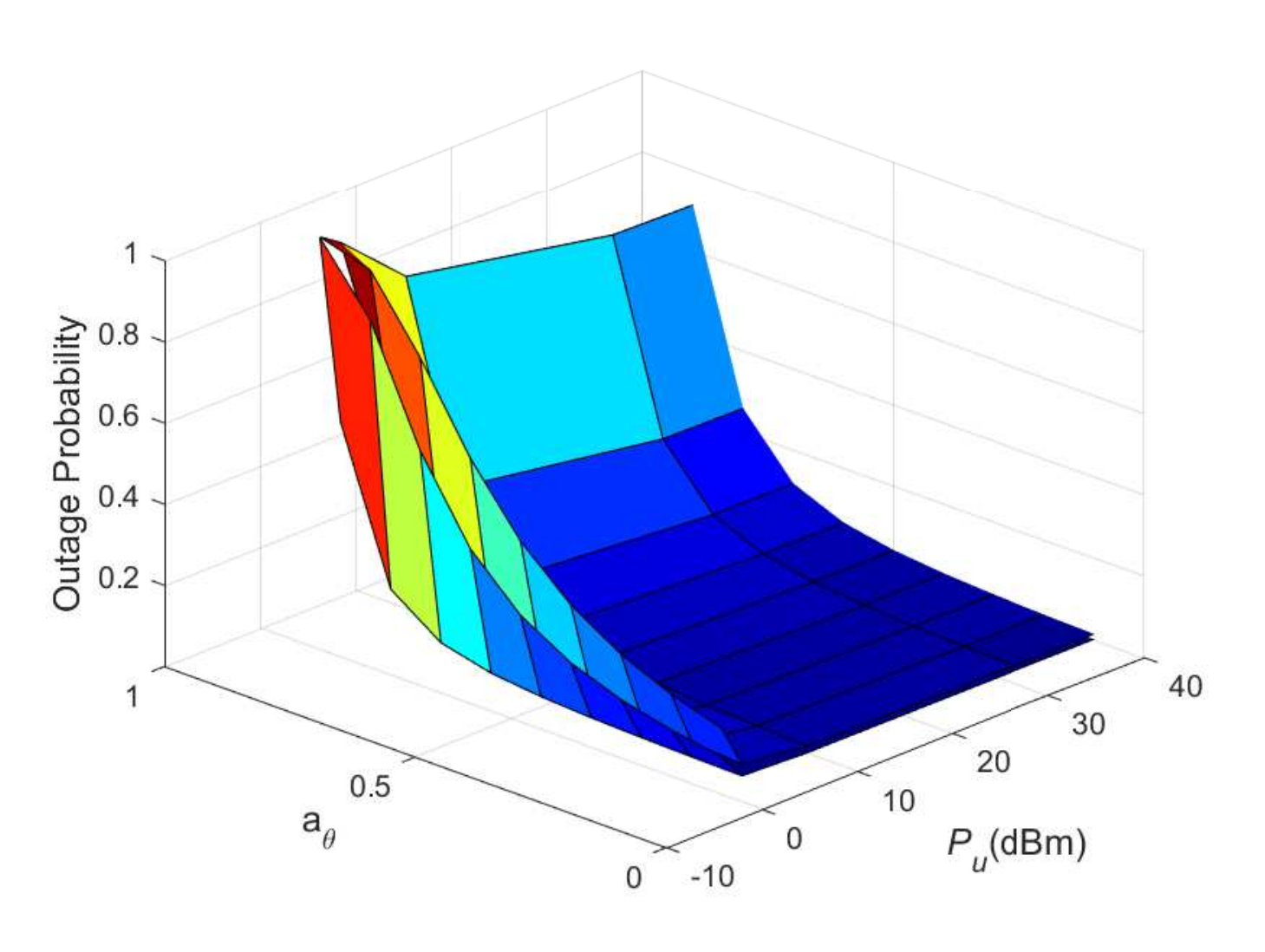}
\caption{ Outage probability versus ${P_u}$ and ${a_\theta}$, with $M = 6$, ${R_1} = 2$ and ${R_2} = 5$ BPCU.}
\label{Fig. 7}
\end{figure}

\begin{figure}[t!]
\centering
\includegraphics[width=3.3in,  height=2.6in]{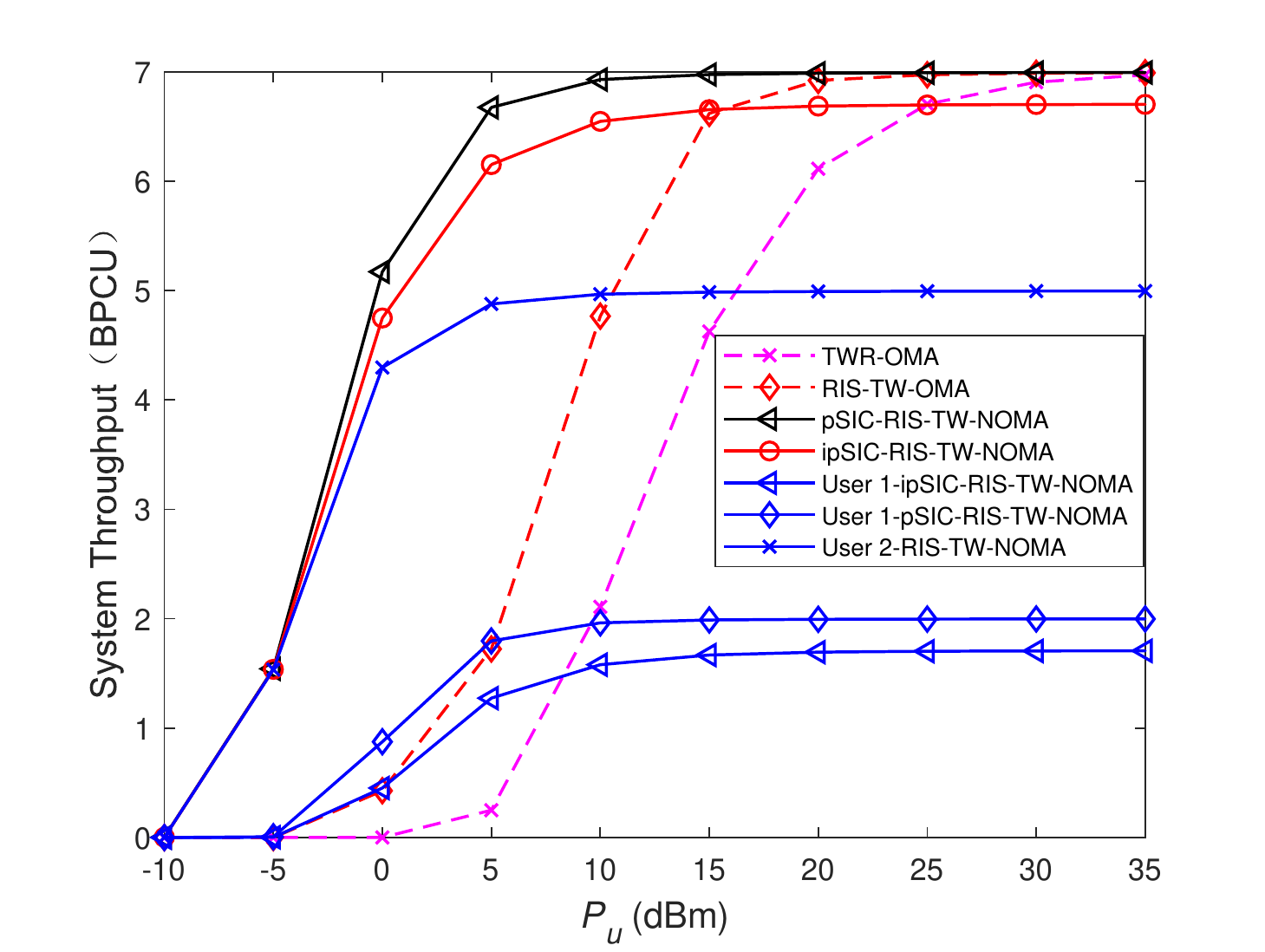}
\caption{System throughput versus ${P_u}$ in delay-limited transmission mode.}
\label{Fig. 8}
\end{figure}

 Fig. \ref{Fig. 4} plots the outage probability of two users versus ${P_u}$ with the different values of target rate for
$\mathbb{E}\left\{ {{{\left| {{g_h}} \right|}^2}} \right\} =  - 6$ dB and
$\mathbb{E}\left\{ {{{\left| {{\sigma _{{I_i}}}} \right|}^2}} \right\} =  - 5$ dB,
and the values of target rate are reduced.
One observation is that the different values of the target rate seriously affect
the outage performance.  As the values of the target rate decrease,
the outage behaviors of users for RIS-TW-NOMA networks become better, followed by the users' behaviors in the conventional OMA networks.
Fig. \ref{Fig. 5} plots the outage probability of two users versus ${P_u}$ with the different values of residual interference
 for ${R_1} = 2$, ${R_2} = 5$ BPCU, $\mathbb{E}\left\{ {{{\left| {{\sigma _{{I_i}}}} \right|}^2}} \right\} =  - 5$ dB and the values of residual interference are $-10$ dB , $-8$ dB and $-5$ dB, respectively. Each red curve corresponds to a residual interference value. The simulated red circle solid curves for outage probability of ${D_1}$ with ipSIC are plotted according to (18), and the exact red fork solid curves for outage probability of ${D_1}$ with ipSIC are plotted according to (19).
It can be seen that the different values of residual interference affect the outage performance seriously. Due to the influence of residual interference from ipSIC,
the outage probability of the nearby user with ipSIC converges to an error floor. As the values of residual interference increase,
the outage behaviors of the ${D_1}$ with ipSIC for RIS-TW-NOMA networks get worse and the preponderance of ipSIC is inexistent compared to pSIC.
Therefore, it is imperative to consider the impact of ipSIC on the RIS-TW-NOMA network's performance in practical applications.

Fig. \ref{Fig. 6} plots the outage probability of users versus ${P_u}$ for a simulation setting with ${R_1} = 2$, ${R_2} = 5$ BPCU, $\mathbb{E}\left\{ {{{\left| {{g_h}} \right|}^2}} \right\} =  - 6$ dB and
$\mathbb{E}\left\{ {{{\left| {{\sigma _{{I_i}}}} \right|}^2}} \right\} =  - 5$ dB.
The approximated outage probability curves for users match precisely with the simulation results.
One can be observed that as the number of RIS elements increases,
the RIS-TW-NOMA network is capable of achieving enhanced outage performance.
The reason is that the application of RIS provides a new
degree of freedom to enhance the wireless link performance.
The conclusions also confirm this phenomenon in \textbf{Remark 2},
where the number of RIS elements influences outage probability for RIS-TW-NOMA.

Fig. \ref{Fig. 7} plots the impact of the power allocation factor ${a_\theta}$ as well as the transmit ${P_u}$ on the performance of the outage probability in the RIS-TW-NOMA scheme, where ${a_\theta}$ is dynamic power allocation coefficients, ${a_\theta } \in \left[ {0,1} \right]$.
We fix a simulation setting with $M = 5$, ${R_1} = 2$, ${R_2} = 5$ BPCU, $\mathbb{E}\left\{ {{{\left| {{\sigma _{{I_i}}}} \right|}^2}} \right\} =  - 5$ dB, and $\mathbb{E}\left\{ {{{\left| {{g_h}} \right|}^2}} \right\} =  - 6$ dB.
Additionally, we assume that the power allocation coefficients of ${D_1}$ and ${D_2}$
 have the relationships of ${a_1} = 1 - {a_\theta }$ and ${a_2} = {a_\theta }$.
In this figure, it is illustrated that with the increase of the transmit ${P_u}$,
the power allocation factor ${a_\theta}$ will be close to 1 when the maximum outage probability is achieved.
In addition, the outage behavior of the nearby user becomes worse, and the performance of the distant user becomes better
with the increase of the power allocation factor ${a_\theta}$.
This is due to the fact that ${D_1}$ suffers more interference from ${D_2}$.
\begin{figure}[t!]
\centering
\includegraphics[width=3.3in,  height=2.6in]{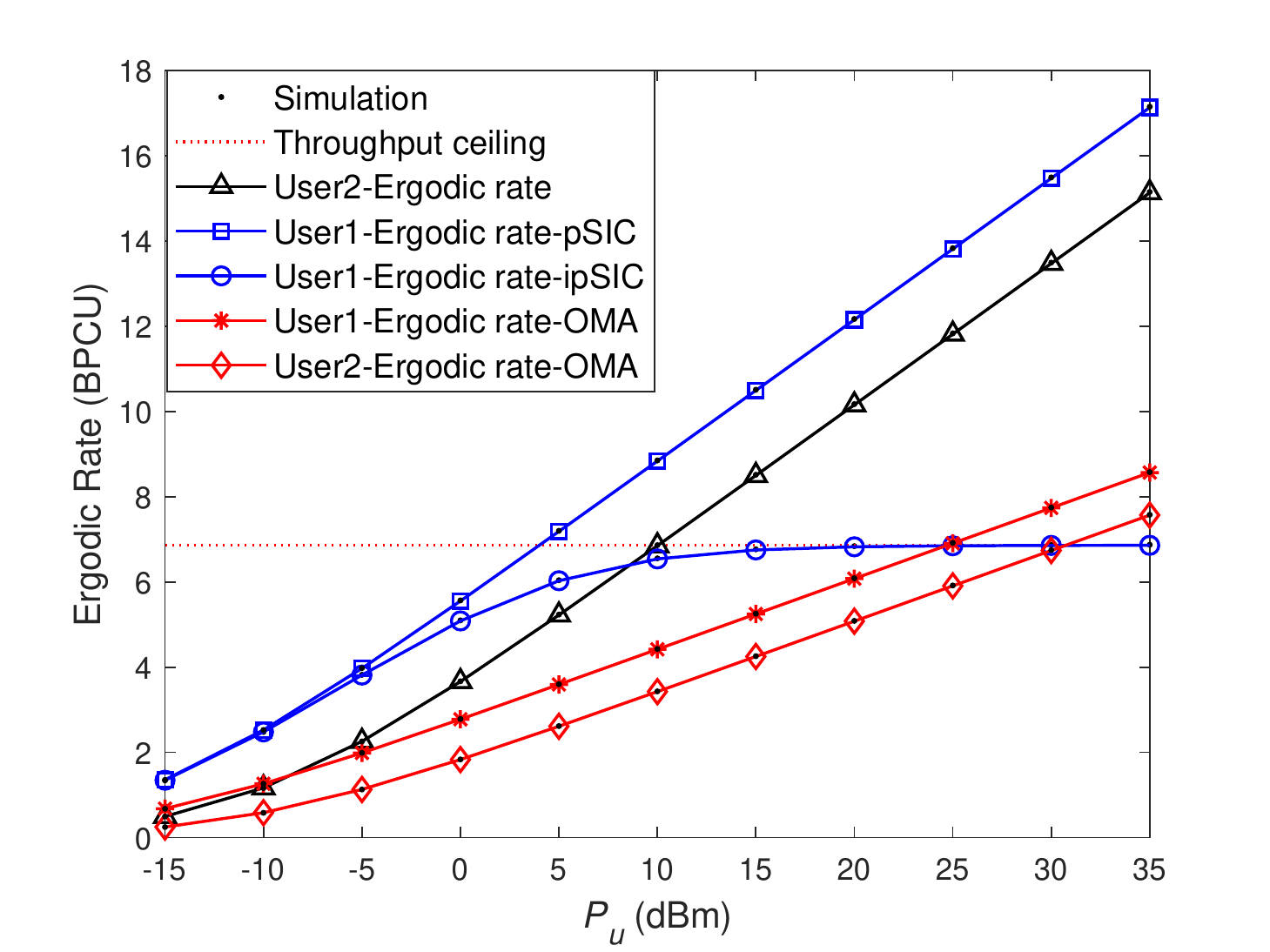}
\caption{ Ergodic rate versus ${P_u}$, with $M = 8$, ${R_1} = 2$, ${R_2} = 5$ BPCU and $\mathbb{E}\left\{ {{{\left| {{g_h}} \right|}^2}} \right\} =  - 6$ dB.}
\label{Fig. 9}
\end{figure}

\begin{figure}[t!]
\centering
\includegraphics[width=3.3in,  height=2.6in]{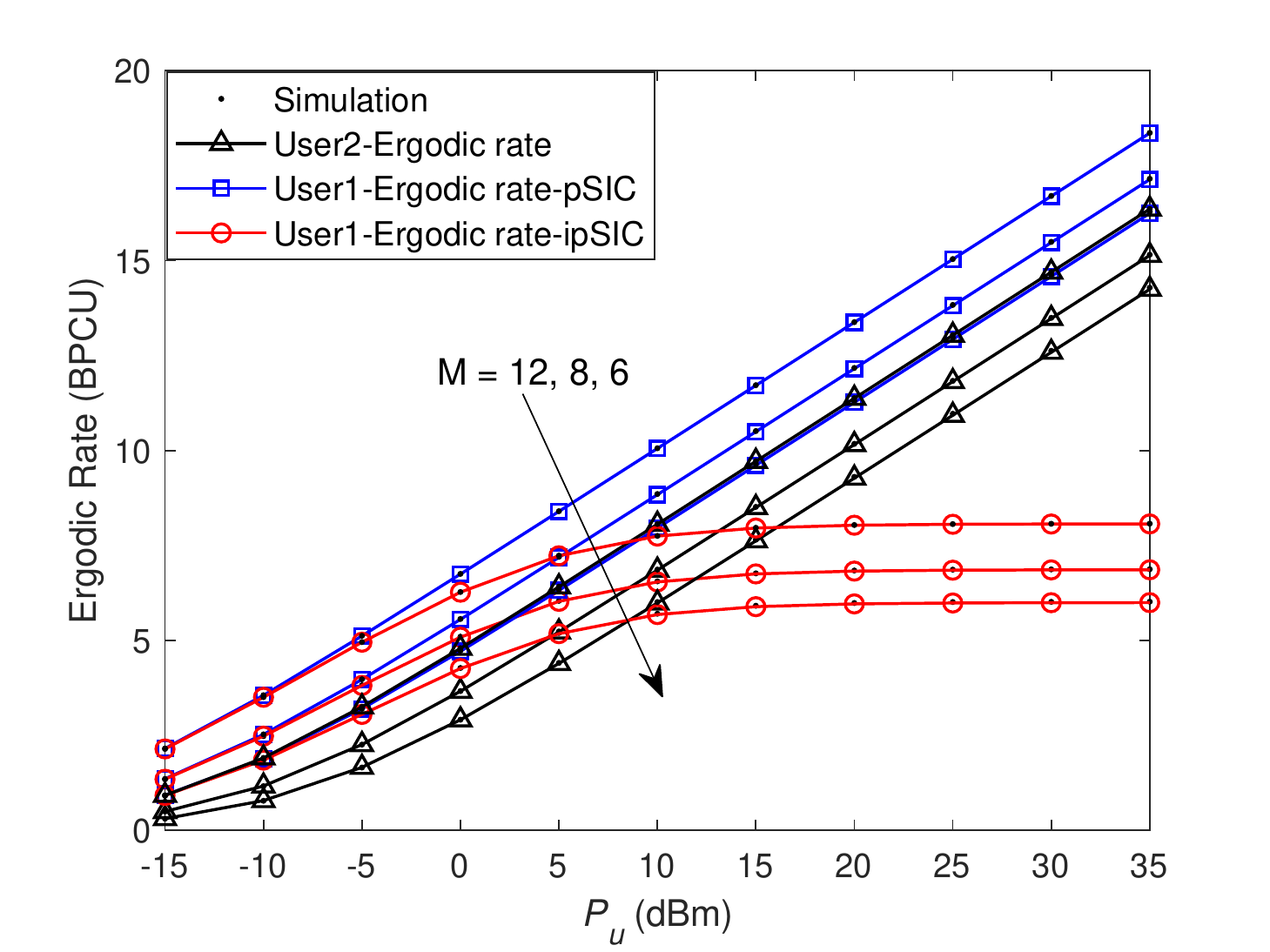}
\caption{ Ergodic rate versus ${P_u}$, with ${R_1} = 2$, ${R_2} = 5$ BPCU and the different \emph{M}.}
\label{Fig. 10}
\end{figure}

Fig. \ref{Fig. 8} plots the curve of system throughput versus ${P_u}$ in delay-limited transmission mode.
The graph is drawn according to (32), and \emph{M} is set to 8.
In this figure, the curves of system throughput with ${P_u}$ in TWR-OMA, RIS-TW-OMA and RIS-TW-NOMA schemes are drawn, respectively.
As can be observed from the figure that the system throughput in RIS-TW-NOMA is better than that of RIS-TW-OMA in the low SNR region.
In addition, by plotting the system throughput of the NOMA system without RIS-aided under the same conditions,
we can conclude that its performance in system throughput is inferior to the system with RIS assistance.

\subsection{Ergodic Rate}

\begin{figure}[t!]
\centering
\includegraphics[width=3.3in,  height=2.6in]{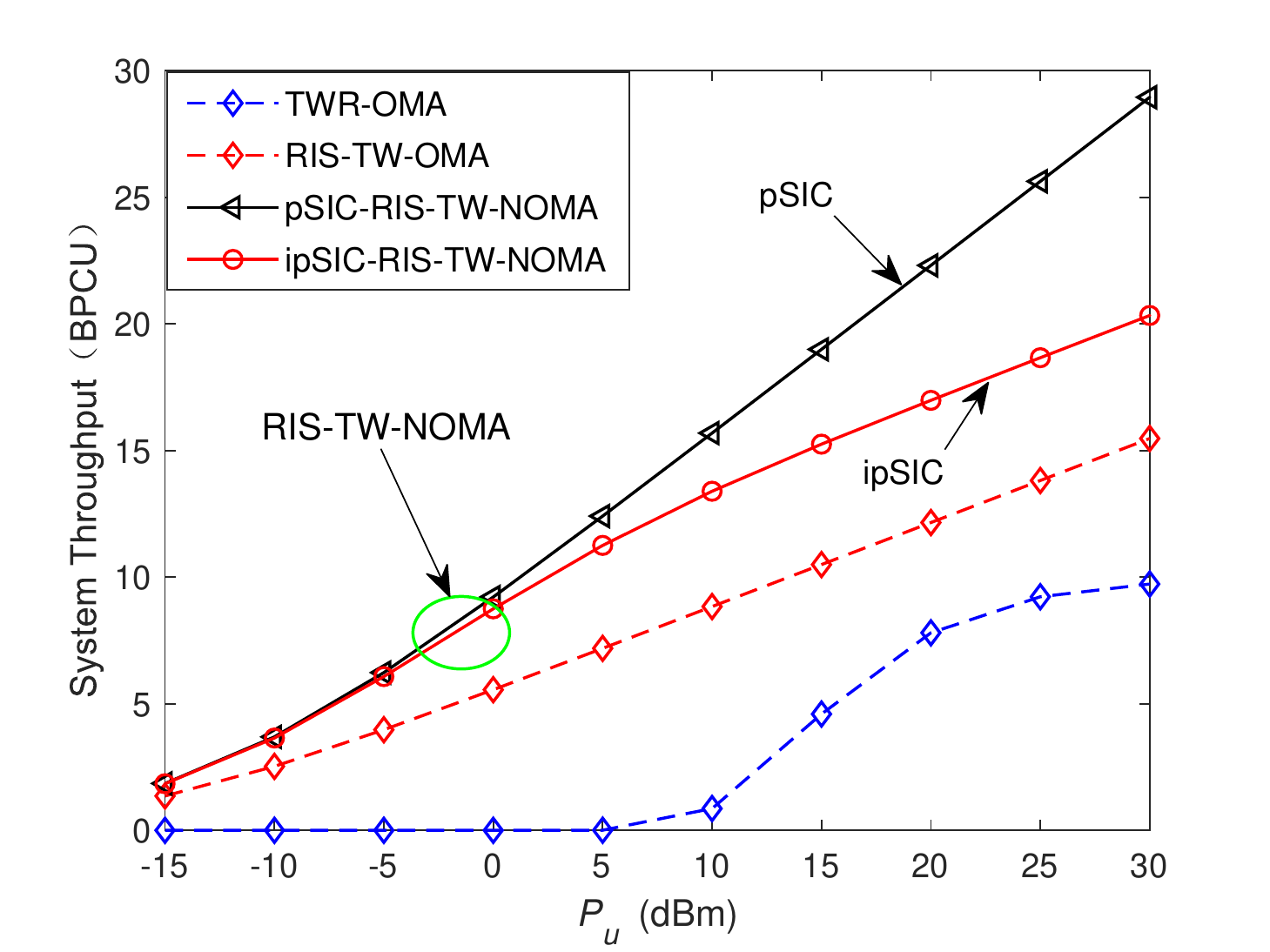}
\caption{System throughput versus ${P_u}$ in delay-tolerant transmission mode.}
\label{Fig. 11}
\end{figure}

Fig. \ref{Fig. 9} plots the ergodic rate of the user versus ${P_u}$ for a simulation setting with ${R_1} = 2$, ${R_2} = 5$ BPCU, $\mathbb{E}\left\{ {{{\left| {{\sigma _{{I_i}}}} \right|}^2}} \right\} =  - 5$ dB, and $\mathbb{E}\left\{ {{{\left| {{g_h}} \right|}^2}} \right\} =  - 6$ dB, which compares the ergodic rates of users in RIS-TW-OMA and RIS-TW-NOMA networks.
These dots are the simulated values, and the lines are the theoretically derived values. In particular, the blue and black solid curves denote the ergodic rates of ${D_2}$ and ${D_1}$ with pSIC/ipSIC for RIS-TW-NOMA networks, which are plotted based on (34), (35) and (37), respectively.
Moreover, the red curve represents the ergodic rates for RIS-TW-OMA networks drawn according to (39) and shown in the figure as a benchmark.
One observation can be drawn that the ergodic rates for RIS-TW-NOMA are much greater than that of RIS-TW-OMA because RIS-TW-NOMA can realize much better user fairness.
The ergodic rates of the nearby user with pSIC and the distant user are better than that of the nearby user with ipSIC.
This reason is that the ergodic rate of the nearby user with ipSIC is affected by residual interference and an ergodic rate ceiling exists.

Fig. \ref{Fig. 10} plots the ergodic rate of users versus ${P_u}$ with the different number of reflecting elements of RIS
for ${R_1} = 2$, ${R_2} = 5$ BPCU, $\mathbb{E}\left\{ {{{\left| {{g_h}} \right|}^2}} \right\} =  - 6$ dB and
$\mathbb{E}\left\{ {{{\left| {{\sigma _{{I_i}}}} \right|}^2}} \right\} =  - 5$ dB.
These dots are the simulated values, and the lines are the theoretically derived values.
One can observe that as the increase for the number of RIS elements, the RIS-TW-NOMA network is capable of achieving enhanced ergodic rate.
Another observation is that the ergodic performance of the nearby user with pSIC has the same slopes with various numbers of RIS elements, which confirms the insights in \textbf{Remark 5}. As a further advance,
Fig. \ref{Fig. 11} plots the curve of system throughput versus ${P_u}$ in delay-tolerant transmission mode
for TWR-OMA, RIS-TW-OMA, and RIS-TW-NOMA.
In this figure, the black and red solid curves represent the system throughput for RIS-TW-NOMA with ipSIC/pSIC, obtained from (43).
The system throughput of TWR-OMA and RIS-TW-OMA are selected to be the benchmarks denoted by the blue and red dash curves.
It is observed that RIS-TW-NOMA can achieve higher throughput, and its tremendous value.
This reason is that the RIS-TW-NOMA network can achieve an enhanced ergodic rate.

\subsection{Energy Efficiency}

 \begin{figure}[t!]
\centering
\includegraphics[width=3.3in,  height=2.6in]{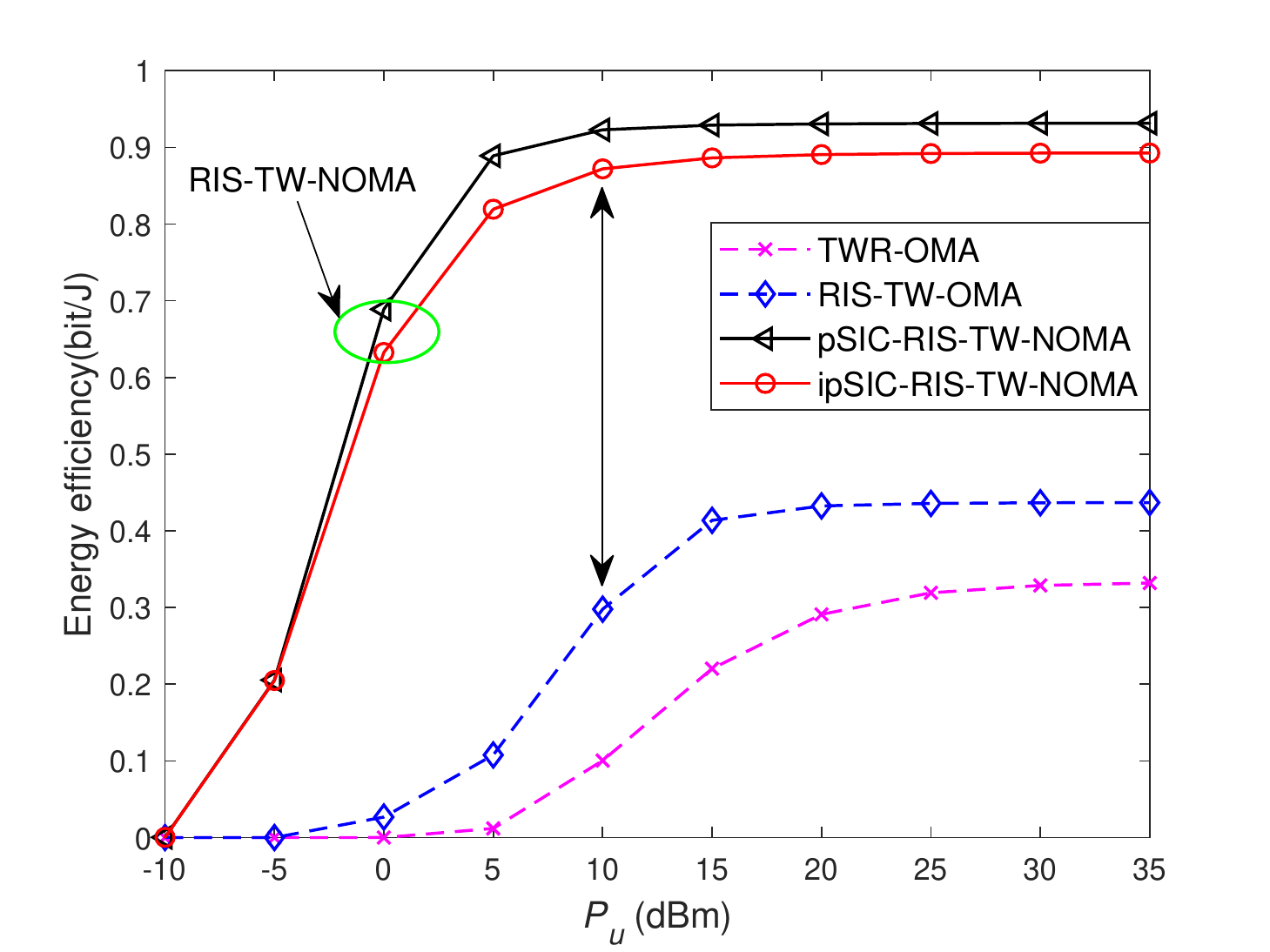}
\caption{Energy efficiency versus ${P_u}$ in delay-limited transmission mode.}
\label{Fig. 12}
\end{figure}

\begin{figure}[t!]
\centering
\includegraphics[width=3.3in,  height=2.6in]{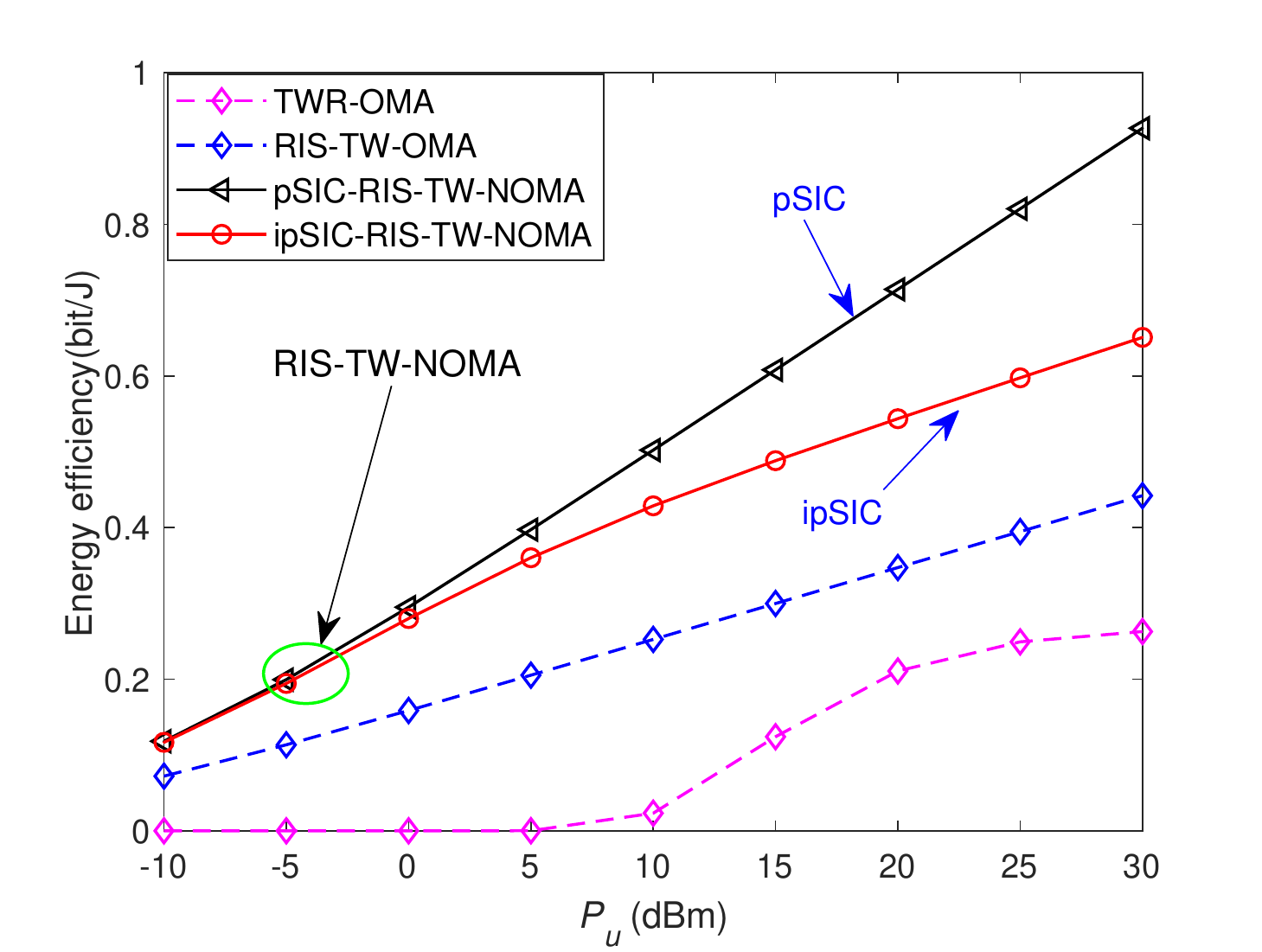}
\caption{Energy efficiency versus ${P_u}$ in delay-tolerant transmission mode.}
\label{Fig. 13}
\end{figure}

 Fig. \ref{Fig. 12} plots the curve of energy efficiency versus ${P_u}$ in delay-limited transmission mode for a simulation setting with ${P} = 1$ dBw,
$\varepsilon  = 1.2$, $ {{P^{{D_1}}}}$ = ${{P^{{D_2}}}} = 10$ dBm,
the energy consumption ${P_{IRS}}$ of IRS hardware is $K{P_k}\left( b \right)$ and ${P_k}\left( b \right) = 10$ dBm.
The energy efficiency curves of RIS-TW-NOMA networks are plotted according to (45).
 It can be observed that RIS-TW-NOMA with ipSIC/pSIC has almost the same energy efficiency in the low SNR region.
 Nevertheless, the energy efficiency of RIS-TW-NOMA with pSIC is superior to ipSIC in the high SNR region.
Another observation is that the energy efficiencies of RIS-TW-NOMA and RIS-TW-OMA networks are superior to that of TWR-OMA networks.
 This is due to the RIS assisted wireless communications are capable of improving energy efficiency compared to these conventional cooperative communications.

Fig. \ref{Fig. 13} plots the curve of energy efficiency for RIS-TW-NOMA in delay-tolerant transmission mode for a simulation setting with ${P} = 1$ dBw,
$\varepsilon  = 2$, $ {{P_1}}$ = ${{P_2}} = 10$ dBm,
the energy consumption ${P_{RIS}}$ of RIS hardware is $K{P_k}\left( b \right)$ and ${P_k}\left( b \right) = 10$ dBm.
The solid curves representing RIS-TW-NOMA with ipSIC/pSIC are obtained from (45),
with throughput in delay-tolerant mode.
 The dashed curves, representing energy efficiency for RIS-TW-OMA and TWR-OMA.
We can observe that the energy efficiency of RIS-TW-NOMA is much larger than that of RIS-TW-OMA and TWR-OMA.
 This is due to that RIS-TW-NOMA can achieve the more significant system throughput relative to these benchmarks.

\section{Conclusion}
This paper has studied the performance of RIS-TW-NOMA networks, where a pair of users are able to exchange their information with the aid of a RIS. The exact and asymptotic expressions of outage probability and ergodic rate for a pair of users with ipSIC/pSIC have been derived. According to the approximated analysis, we derive the diversity orders and high SNR slopes of users for RIS-TW-NOMA networks.
It has been shown that users' outage behaviors and ergodic rates are related to the number of RIS elements.
Additionally, the system throughput and energy efficiency of RIS-TW-NOMA have been discussed in both delay-limited and delay-tolerant transmission modes.
Numerical results have shown that the proposed RIS-TW-NOMA is able to improve the outage performance and ergodic rate compared to RIS-TW-OMA.
 The setting of single antenna and coherent phase shifting may give rise to overestimated performance for RIS-TW-NOMA networks, hence our future work will relax these assumptions. Another promising future research direction is extend to multiple pairs of user research and consider using a random phase shifting approach to deal with the cascade channels, which will further enhance the performance of RIS-TW-NOMA networks.

\section*{Appendix~A: Proof of Theorem \ref{theorem:1}} \label{Appendix:A}
\renewcommand{\theequation}{A.\arabic{equation}}
\setcounter{equation}{0}
According to the definition of outage probability, the outage probability of ${D_1}$ with ipSIC can be expressed by
\begin{align}\label{express20}
P_{{D_1}}^{ipSIC} = {{\rm{P}}{\rm{r}}}[{\gamma _{{D_1} \to {x_2}}} < {\gamma _{t{h_2}}}].
\end{align}
Upon substituting (2) into (18) the outage probability of ${D_1}$ with ipSIC can be further given by
\begin{align}\label{express22}
P_{{D_1}}^{ipSIC}
&= {{\rm{P}}{\rm{r}}}\left[ {\frac{{{P_u}{a_2}{{\left| {{{\bf{h}}^{H}}{\bf {\Phi}}  {\bf{g}}} \right|}^2}}}{{\varepsilon {P_u}{{\left| {{g_h}} \right|}^2} + \sigma _{{I_1}}^2{\rm{ + }}\sigma _{{n_1}}^2}} < {\gamma _{t{h_2}}}} \right]\nonumber\\
&= {{\rm{P}}{\rm{r}}}\left[ {{{\left| {{{\bf{h}}^{H}}{\bf {\Phi}}  {\bf{g}}} \right|}^2} < \frac{{{\gamma _{t{h_2}}}}}{{{P_u}{a_2}}}(\varepsilon {P_u}\sigma _{{g_h}}^2 + \sigma _{{I_1}}^2{\rm{ + }}\sigma _{{n_1}}^2)} \right],
\end{align}
where $\chi  = \left| {{{\bf{h}}^H}{\bf {\Phi}}  {\bf{g}}} \right|{\rm{ = }}\left| {\sum\limits_{m = 1}^M {{h_m}{g_m}{e^{ - j{\theta _m}}}} } \right|$. For the coherent phase shifting design, the phase shifts of the RIS are matched with the phases of the RIS fading gains, the $\chi$ can be further expressed as
$\sum\limits_{m = 1}^M {\left| {{h_m}{g_m}} \right|}$.

By using the CLT, $\frac{{\chi  - M{\mu _{\left| {{h_m}{g_m}} \right|}}}}{{\sqrt M {\sigma _{\left| {{h_m}{g_m}} \right|}}}}$ obeys the standard normal distribution ${\cal N}\left( {0,1} \right)$. $\chi$ can be approximated as the following Gaussian random variable:
\begin{align}\label{express19}
\chi  = \left| {\sum\limits_{m = 1}^M {{h_m}{g_m}} } \right| \sim  {\cal N}( {M{\mu _{\left| {{h_m}{g_m}} \right|}},M\sigma _{\left| {{h_m}{g_m}} \right|}^2} ).
\end{align}
And, $X$ can be approximated as the following Gaussian random variable:
\begin{align}\label{express19}
X = \sqrt M \left( {\frac{\chi }{M} - {\mu _{_{\left| {{h_m}{g_m}} \right|}}}} \right)\sim{\cal N}( {0,\sigma _{_{_{\left| {{h_m}{g_m}} \right|}}}^2}).
\end{align}
As a result, the outage probability of ${D_1}$ with ipSIC can be approximated as
\begin{align}\label{express19}
P_{{D_1}}^{ipSIC}
&=  {{\rm{P}}{\rm{r}}}\left[ {{\chi ^2} < {\tau ^2}} \right] \nonumber\\
&= {{\rm{P}}{\rm{r}}}\left[ {X < \sqrt M \left( {\frac{\tau }{M} - {\mu _{\left| {{h_m}{g_m}} \right|}}} \right)} \right],
\end{align}
where $\tau  = \sqrt {\frac{{{\gamma _{t{h_2}}}}}{{{P_u}{a_2}}}(\varepsilon {P_u}\sigma _{{g_h}}^2 + \sigma _{{I_1}}^2{\rm{ + }}\sigma _{{n_1}}^2)} $.

Applying the some algebraic manipulations, the closed-form expression for the outage probability of ${D_1}$ with ipSIC in RIS-TW-NOMA networks is calculated as
\begin{align}\label{express22}
P_{{D_1}}^{ipSIC}
&\approx \frac{1}{2} \!+\!\! \int_0^{\sqrt M \left( {\frac{\tau }{M} - {\mu _{\left| {{h_m}{g_m}} \right|}}} \right)} {\frac{1}{{\sqrt {2\pi } {\sigma _{\left| {{h_m}{g_m}} \right|}}}}} {e^{ - \frac{{{t^2}}}{{2\sigma _{\left| {{h_m}{g_m}} \right|}^2}}}}dt\nonumber\\
&= \frac{1}{2} + \frac{1}{{\sqrt \pi  }}\phi \left[ {\frac{{\sqrt M \left( {\frac{\tau }{M} - {\mu _{_{\left| {{h_m}{g_m}} \right|}}}} \right)}}{{\sqrt 2 {\sigma _{_{\left| {{h_m}{g_m}} \right|}}}}}} \right].
\end{align}
Upon substituting (14) and (15) into (A.6), the closed-form expression for the outage probability of ${D_1}$ with ipSIC in RIS-TW-NOMA networks can be given by
\begin{align}\label{express22}
P_{{D_1}}^{ipSIC}= \frac{1}{2} + \frac{1}{{\sqrt \pi  }}\phi \left[ {\sqrt {\frac{M}{{2(1 - \frac{{{\pi ^2}}}{{16}})}}} \left( {\frac{\tau }{M} - \frac{\pi }{4}} \right)} \right],
\end{align}
where $\phi \left( x \right) \buildrel \Delta \over = \int_0^x {{e^{ - {t^2}}}} dt $.
The proof is completed.

\section*{Appendix~B: Proof of Theorem \ref{theorem:2}} \label{Appendix:B}
\renewcommand{\theequation}{B.\arabic{equation}}
\setcounter{equation}{0}
The inclusion of Bessel functions in the pdf of $\left| {{h_m}{g_m}} \right|$ is the main reason for the difficulty of performance analysis.
Based on ~\cite{yang2017approximating}, the upper bound of the Bessel function was provided by
\begin{align}\label{express22}
{K_0}\left( x \right) \le \sqrt {\frac{\pi }{{2x}}} {e^{ - x}}.
\end{align}

With the aid of the upper bound for the Bessel function, the upper bound for PDF of $\chi  = \left| {\sum\limits_{m = 1}^M {{h_m}{g_m}} } \right|$
can be calculated as follows
\begin{align}\label{express22}
{f_{\left| {{h_m}{g_m}} \right|}}(x) &= 4x{K_0}\left( {2x} \right) \le 4x\sqrt {\frac{\pi }{{4x}}} {e^{ - 2x}} \nonumber\\
&= 2\sqrt {\pi x} {e^{ - 2x}} \buildrel \Delta \over = \upsilon \left( x \right).
\end{align}

 Because of the expression of $\upsilon \left( x \right)$, an upper bound of $\sum\nolimits_{i = 1}^M {{x_i}}$ is derived as follows and
 the Laplace transform of $\upsilon \left( x \right)$ is written as
\begin{align}\label{express22}
{\cal L}(\upsilon (x))
&= \int_0^\infty  2 \sqrt {\pi x} {e^{ - 2x}}{e^{ - sx}}dx\nonumber\\
&= 2\sqrt \pi  \int_0^\infty  {{x^{\frac{1}{2}}}} {e^{ - 2x}}{e^{ - sx}}dx= \frac{{2\sqrt \pi  \Gamma \left( {\frac{3}{2}} \right)}}{{{{\left( {s + 2} \right)}^{\frac{3}{2}}}}}.
\end{align}
where $\Gamma (x) = \int_0^{ + \infty } {{t^{x - 1}}} {e^{ - t}}dt\left( {x > 0} \right)$ is the Gamma function.

Due to the fact that ${x_i}$ is independent and identically distributed, the pdf of the sum is denoted by ${f_{\sum\nolimits_{i = 1}^M {{x_i}} }}(x)$,
 thus, ${f_{\sum\nolimits_{i = 1}^M {{x_i}} }}(x)$ can be upper bounded as
\begin{align}\label{express22}
{f_{\sum\nolimits_{i = 1}^M {{x_i}} }}(x)& \le{{\cal L}^{{\rm{ - }}1}}\left( {{{\left( {{\cal L}(\upsilon (x))} \right)}^M}} \right)\nonumber\\
&={{\cal L}^{{\rm{ - }}1}}\left( {\frac{{{2^M}{\pi ^{\frac{M}{2}}}{\Gamma ^M}\left( {\frac{3}{2}} \right)}}{{{{\left( {s + 2} \right)}^{\frac{3}{2}M}}}}} \right).
\end{align}
Assume that $M$ is an even number and make $L= \frac{M}{2}$, the upper bound of ${f_{\sum\nolimits_{i = 1}^M {{x_i}} }}(x)$ can be given by
\begin{align}\label{express22}
{f_{\sum\nolimits_{i = 1}^M {{x_i}} }}(x) \le \frac{{{2^M}{\pi ^{\frac{M}{2}}}{\Gamma ^M}\left( {\frac{3}{2}} \right)}}{{(3L   - 1)!}}{x^{3L   - 1}}{e^{ - 2x}}.
\end{align}

Based on (18), The upper bound for outage probability of ${D_1}$ with ipSIC in RIS-TW-NOMA networks can be given by
\begin{align}\label{express22}
P_{{D_1}}^{ipSIC}
&= {{\rm{P}}{\rm{r}}}\left[ {{{\left| {{{\bf{h}}^{H}}{\bf {\Phi}}  {\bf{g}}} \right|}^2} < {\tau  ^2}} \right] = {{\rm{P}}{\rm{r}}}\left[ {\sum\limits_{m = 1}^M {\left| {{h_m}{g_m}} \right|}  < \tau } \right]\nonumber\\
&\le \int_0^\tau  {\frac{{{2^M}{\pi ^{\frac{M}{2}}}{\Gamma ^M}\left( {\frac{3}{2}} \right)}}{{(3L - 1)!}}{x^{3L   - 1}}{e^{ - 2x}}dx}\nonumber\\
&= \frac{{{2^{ - L}}{\pi ^{\frac{M}{2}}}{\Gamma ^M}\left( {\frac{3}{2}} \right)}}{{(3L - 1)!}}\gamma \left( {3L,2\tau } \right),
\end{align}
where $\phi \left( x \right) \buildrel \Delta \over = \int_0^x {{e^{ - {t^2}}}} dt $,
 $\tau  = \sqrt {\frac{{{\gamma _{t{h_2}}}}}{{{P_u}{a_2}}}(\varepsilon {P_u}\sigma _{{g_h}}^2 + \sigma _{{I_1}}^2{\rm{ + }}\sigma _{{n_1}}^2)} $, $\gamma (s,x) = \int_0^x {{t^{s - 1}}} {{\rm{e}}^{ - t}}dt$ is the incomplete Gamma function.
The proof is completed.

\section*{Appendix~C: Proof of Theorem \ref{theorem:3}} \label{Appendix:C}
\renewcommand{\theequation}{C.\arabic{equation}}
\setcounter{equation}{0}

Upon substituting ${\varepsilon= 1}$ into (33), the ergodic rate of ${D_1}$ with ipSIC in RIS-TW-NOMA networks can be written as
\begin{align}\label{express30}
R_{{D_1}}^{ipSIC}
&=\mathbb{E} \left[ {\log \left( {1 + \underbrace {\frac{{{P_u}{a_2}{{\left| {{{\bf{h}}^{H}}{\bf {\Phi}}  {\bf{g}}} \right|}^2}}}{{{P_u}{{\left| {{g_h}} \right|}^2} + \sigma _{{I_1}}^2{\rm{ + }}\sigma _{{n_1}}^2}}}_Y} \right)} \right]\nonumber\\
&= \frac{1}{{\ln 2}}\int_0^\infty  {\frac{{1 - {F_Y}(y)}}{{1 + y}}}dy,
\end{align}
where $Y = \frac{{{P_u}{a_2}{{\left| {{{\bf{h}}^{H}}{\bf {\Phi}}  {\bf{g}}} \right|}^2}}}{{\varepsilon {P_u}{{\left| {{g_h}} \right|}^2} + \sigma _{{I_1}}^2{\rm{ + }}\sigma _{{n_1}}^2}}$.

The CDF of Y can be expressed as
\begin{align}\label{express30}
{F_Y}\left( y \right) = {\rm{Pr}}\left( {\frac{{{P_u}{a_2}{{\left| {{{\bf{h}}^{H}}{\bf {\Phi}}  {\bf{g}}} \right|}^2}}}{{\varepsilon {P_u}{{\left| {{g_h}} \right|}^2} + \sigma _{{I_1}}^2{\rm{ + }}\sigma _{{n_1}}^2}} < y} \right).
\end{align}

With the aid of (17), the CDF of Y can be calculated as
\begin{align}\label{express30}
{F_Y}(y)
&= {{\rm{P}}{\rm{r}}}\left[ {{{\left| {{{\bf{h}}^{H}}{\bf {\Phi}}  {\bf{g}}} \right|}^2} < \frac{y}{{{P_u}{a_2}}}(\varepsilon {P_u}\sigma _{{g_h}}^2 + \sigma _{{I_1}}^2{\rm{ + }}\sigma _{{n_1}}^2)} \right]\nonumber\\
&\approx \frac{1}{2} + \int_0^\kappa  {\frac{1}{{\sqrt {2\pi } {\sigma _{\left| {{h_m}{g_m}} \right|}}}}} {e^{ - \frac{{{t^2}}}{{2\sigma _{\left| {{h_m}{g_m}} \right|}^2}}}}dt,
\end{align}
where $\kappa  = \sqrt M \left( {\sqrt {\frac{{y(\varepsilon {P_u}\sigma _{{g_h}}^2 + \sigma _{{I_1}}^2{\rm{ + }}\sigma _{{n_1}}^2)}}{{{P_u}{a_2}}}} \frac{1}{M} - {\mu _{\left| {{h_m}{g_m}} \right|}}} \right)$.

Upon substituting (14) and (15) into (C.3), the CDF of Y is given by
\begin{align}\label{1}
\begin{split}
{F_Y}\left( y \right)
&= \frac{1}{2} + \frac{1}{{\sqrt \pi  }}\phi \left[ {\sqrt {\frac{M}{{2(1 - \frac{{{\pi ^2}}}{{16}})}}} } \right.\\
&\left. { \times \left( {\sqrt {\frac{{y(\varepsilon {P_u}\sigma _{{g_h}}^2 +\sigma _{{I_1}}^2{\rm{ + }}\sigma _{{n_1}}^2)}}{{{P_u}{a_2}}}} \frac{1}{M} - \frac{\pi }{4}} \right)} \right].
\end{split}
\end{align}

Combining (C.1) and (C.4), the exact expression for ergodic rate of ${D_1}$ with ipSIC in RIS-TW-NOMA networks can be given by
\begin{align}\label{1}
\begin{split}
R_{{D_1}}^{ipSIC} = & \frac{1}{{\ln 2}}\int_0^\infty  {\frac{1}{{1 + y}}} \left\{ {\frac{1}{2} - \frac{1}{{\sqrt \pi  }}\phi \left[ {\sqrt {\frac{M}{{2(1 - \frac{{{\pi ^2}}}{{16}})}}} } \right.} \right.\\
&\left. {\left. { \times \left( {\sqrt {\frac{{y(\varepsilon {P_u}\sigma _{{g_h}}^2 + \sigma _{{I_1}}^2{\rm{ + }}\sigma _{{n_1}}^2)}}{{{P_u}{a_2}}}} \frac{1}{M} - \frac{\pi }{4}} \right)} \right]} \right\}dy.
\end{split}
\end{align}
The proof is completed.

\bibliographystyle{IEEEtran}
\bibliography{mybib}

\end{document}